\documentclass[pra,twocolumn,superscriptaddress]{revtex4}%
\usepackage{amssymb}
\usepackage{amsmath}
\usepackage{graphicx}
\usepackage{dcolumn}
\usepackage{color}
\usepackage{bm}
\usepackage{subfigure}
\usepackage{amsfonts}
\usepackage{appendix}
\usepackage{xcolor}%
\providecommand{\U}[1]{\protect\rule{.1in}{.1in}}

\begin{document}
\preprint{APS/123-QED}
\title{Approximate Quantum Algorithms as a Multiphoton Raman Excitation of a
Quasicontinuum Edge}
\author{A.\ Mandilara}
\affiliation{Department of Informatics and Telecommunications, National and Kapodistrian
University of Athens, Panepistimiopolis, Ilisia, 15784, Greece}
\author{Daniil Fedotov}
\affiliation{Department of Electrical Engneering, Higher School of Economics, 34,
Tallinskaya st., Moscow, 123458, Russia.(on leave)}
\author{V.\ M.\ Akulin}
\affiliation{Laboratoire Aim\'{e}-Cotton CNRS UMR 9025, L'Universit\'{e} Paris-Saclay,
B\^{a}t. 505, Campus d'Orsay, Rue Aim\'{e}-Cotton, 91405 Orsay Cedex, France.}
\affiliation{Institute for Information Transmission Problems of the Russian Academy of
Science, Bolshoy Karetny per. 19, Moscow, 127994, Russia.(on leave)}

\begin{abstract}
Many quantum algorithms can be seen as a transition from a well-defined
initial quantum state of a complex quantum system, to an unknown target
quantum state, corresponding to a certain eigenvalue either of the Hamiltonian
or of a transition operator. Often such a target state corresponds to the
minimum energy of a band of states. In this context, approximate quantum
calculations imply transition not to the single, minimum energy, state but to
a group of states close to the minimum. We consider dynamics and the result of
two possible realization of such a process -- transition of population from a
single initially populated isolated level to the quantum states at the edge of
a band of levels. The first case deals with the time-independent Hamiltonian,
while the other with a moving isolated level. We demonstrate that the energy
width of the population energy distribution over the band is mainly dictated by
the time-energy uncertainty principle, although the specific shape of the
distribution depends on the particular setting. We consider the role of the
statistics of the coupling matrix elements between the isolated level and the
band levels. We have chosen the multiphoton Raman absorption by an ensemble of
Rydberg atoms as the model for our analysis, although the results obtained can
equally be applied to other quantum computing platforms.

\end{abstract}
\maketitle

\section{ \ \bigskip The context of the problem}

Interaction of an isolated level with a large number of other states is the
well-known classical problem that is often encountered in different domains of
Quantum Physics and Physical Chemistry. Active research in the field of
Quantum Computing during the last decade brought to the fore an important
facet of this problem -- how to efficiently populate states of a continuous
spectrum edge. In this paper we consider a number of possible regimes of the
population dynamics taking place at a quasi-continuum edge in the course of it
interaction-induced excitation.

The problem of a level interacting with a band has a very long history, which
starts from the Fermi's golden rule and has its continuation in many other
publications revealing various aspects of the process, such as the population
distribution over the energy scale considered by Jortner and
Bixon \cite{Jortner Biggs}, the interference induced by the presence of a
strong coupled level known as Fano profile \cite{Fano}, the role of the level
motion \ considered by Demkov and Osherov \cite{DemkovOcherov}, and the role
of the level couplings statistics \cite{Akulin}, among other papers in this
field. All these aspects are also important at the spectrum edge, affecting
the rate at which the transition occurs and the energy width of the population distribution.

A first indication of the underlying relation between level-band problems and
quantum computing was given with the introduction of the adiabatic quantum
computation model \cite{Farhi}. In this model the solution to a classical
problem is mapped into the ground state of the spectrum of a many-body
Hamiltonian and the role of the quantum algorithm is to prescribe the
adiabatic procedure for reaching the latter. Soon it became evident that for
the most interesting combinatorial problems \cite{Lidar} the required, for the
success of the adiabatic procedure, gap between the ground state and the rest
of the spectrum, is not guaranteed. As a result new algorithmic procedures
\cite{Farhi2} appeared with the aim to offer approximate solutions to such
optimization problems and which can potentially give results even in the absence of
a gap. These approximate methods having also the advantage of requiring less
physical resources, have attracted the interest of the community during the
last years. In short, the generic prescription for reaching the edge of the
spectrum is to adjust the parameters, such as application times of fixed
control signals, in accordance with classically processed feedback from the
quantum system. The results of the current work, also provide guidance for
populating quantum states at the edge of the spectrum although the procedure
under consideration is designed for a specific quantum `hardware', i.e.,
arrays of Rydberg atoms, exploiting long-standing theoretical and experimental
knowledge on such physical systems. Recent experiments \cite{Lukin2}
demonstrate the actuality of such a choice.

It is worth mentioning that the straightforward dipole-dipole mechanism is not
the only possible way to create interaction among atoms. A more general case
of interaction can be constructed by periodic change of the positions of
interacting atoms, which can be seen as periodic sequence of unitary
operations applied to single atoms, pairs of them, and to groups of a few
atoms. Quantum evolution in this case can be interpreted as action of an
"effective" Hamiltonian given by the logarithm of the unitary transformation
over the period. In the presence of an external electromagnetic field, one can
therefore speak about the controlled multi-quantum transitions among the
eigenstates of this Hamiltonian, the quasienergy levels of the atomic ensemble
under periodic manipulations. The aim remains the same -- to populate the
lowest level of the quasienergy band, if we speak about the exact algorithm,
or to transfer population to a narrow strip of the levels nearest to the edge,
when admitting approximate algorithms. Coherence time for such a setting is
dictated by the accuracy with which the periodic operations are performed.

The paper is organized as follows. We start by presenting a well-known
physical model of quantum computation based on an ensemble of Rydberg
atoms interacting with an electromagnetic field in the regime of the
dipole-dipole blockade \cite{Lukin} and discuss how it can be seen in terms of
the multiphoton transitions. We then remind the main features of the
level-band system dynamics employing qualitative images that help to better
perceive different excitation regimes. After these introductory sections, we
turn to the main part of the paper by considering the excitation of a uniform
continuum edge after an abrupt switch-on interaction, and calculate the energy
distribution of the population and the required time it to attain. At a
further step, we discuss the possibility of narrowing the distribution by
slowly approaching the level from the infinitely far energy position to the
edge and back and discuss the regime where the distribution width is limited
just be the frequency-time uncertainty principle. Next question we address is
the role of the distribution of the coupling energies between the level and
the states of the continuum. We show that the continuum inhomogeneity is
capable of ``spoiling'' the energy distribution under certain conditions. We
conclude by considering a rather general case of the coupling statistics and
identify the regime where a rather narrow population distribution at the
continuum edge can be achieved in spite of the presence of the ``spoiling
states''. We finally discuss what the  obtained results mean for the approximate
quantum algorithms.

\section{The model}

There are many possible physical realization of two-level models suitable for
Quantum Calculations. We focus on one of them -- an ensemble of two-level
Rydberg atoms, where both of levels correspond to the highly excited Rydberg
states coupled to the external intensive and highly coherent RF field by
dipole interaction. The RF field itself can be either monochromatic, such that
the field frequency is close to the frequency of the transition between the
Rydberg states, or contain more harmonics, such that the transition frequency
and the field frequencies are close to the condition of a multiphoton Raman
resonance. The long wavelength of the RF radiation implies that all the
ensemble atoms are placed in the same external field. At the same time,
periodic manipulations exert upon the atomic position by the technique of
optical tweezers allows one to realize the effective multiparticle
interactions among atoms in addition to their regular second-order van der
Waals dipole interactions. The lower Rydberg state will hereafter be referred
as the ground state, while the upper Rydberg state as "excited".

There are also many possible formulation of the mathematical problem which is
supposed to be solved by physical realization of a Quantum Algorithm. The most
known formulation requires population of the lowest energy eigenstate, or at
least the eigenstates close in energy to the minimum. Here we focus on a
slightly different formulation of the problem and require population of the
state corresponding to the minimum energy per excited atom. The multiphoton
Raman excitation enable one to select from the entire variety of such states
some submanifolds satisfying the Raman resonance condition. In a sense, it is
a more general formulation of the mathematical problem, which however is
equivalent to the most known problem from the point of view of the complexity theory.

In order to be more specific, let us consider the Hamiltonian of such an
ensemble of \ $N$ two-level atoms without RF field in the form
\begin{equation}
\widehat{H}=\sum_{n=1}^{N}\hslash\omega\widehat{\sigma}_{z,n}+F\left(
\left\{  \widehat{\sigma}_{z,n}\right\}  \right)  , \label{Eq1}%
\end{equation}
which consist of the sum of individual atomic Hamiltonian $\hslash
\omega\widehat{\sigma}_{z,n}$ and a nonlinear function $F\left(  \left\{
\widehat{\sigma}_{z,n}\right\}  \right)  $ given in terms of the Pauli
operators $\hslash\omega\widehat{\sigma}_{z,n}$. The nonlinear function is
constructed with the help of the second order van der Waals coupling and the
periodic control technique in such a way, that it takes the minimum value per
exited atom for a quantum state of the ensemble corresponding to the solution
of a mathematical problem of interest.

The Hamiltonian Eq.(\ref{Eq1}) is diagonal in the basis of direct products of
the lower and upper states of the atoms. This is the so-called "computational"
basis, where each of its eigenstate has the form%
\[
\left\vert 0,1,\ldots,1,0\right\rangle
\]
and corresponds to the binary representation of an integer, encoded in the
distribution of the excitation of different atoms whence each atom corresponds
to a certain register of this binary encoding. Position of the lowest
eigenvalue is supposed to be known. What is required is to find the
eigenvector corresponding to this lowest eigenvalue in the computational
basis. Our aim is therefore to transfer all the population from the ground
state of the atomic ensemble to this very state, or at least, to a group of
states close to the minimum energy per atom. Once this goal achieved, by
measuring the excitations distribution among individual atoms, one finds the
required integer -- either the exact solution of the mathematical problem
encoded in the Hamiltonian $\widehat{H}$ via the nonlinear function $F\left(
\left\{  \widehat{\sigma}_{z,n}\right\}  \right)  $, or an approximate
solution -- another number, which gives the energy value close to the minimum.

\begin{figure}
[h]
\begin{center}
\includegraphics[
height=3.3451in,
width=1.7227in
]%
{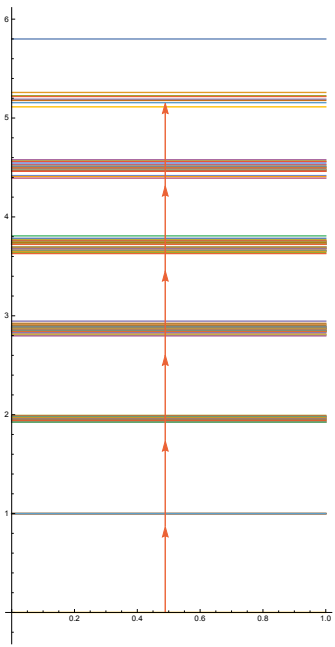}%
\caption{The energy spectrum $E_{s}$ of seven two-level atoms interacting via
their $\widehat{\sigma}_{z}$. Parameter of two,three, and for-particle
interactions are chosen randomly. This spectrum mimics a spectrum of
Hamiltonian constructed for finding a minimum energy per atom for a specific
problem. By arrows we depict energies corresponding to the absorption of a
number of photons. For the chosen parameters, the minimum energy per atoms
corresponds to the excitation of six atoms.}%
\label{Fig1a}%
\end{center}
\end{figure}

The Hamiltonians are realized by constructing the interatomic interactions.
One may think of finding the minimum energy for the Ising problem, as an
example. A more general example of the spectrum of an ensemble of seven atoms
generated by a random choice of constants of the binary, triple, and four-body
interatomic interactions is depicted in Fig.\ref{Fig1a}.%

As a possible physical realization of such an ensemble, one may think of a
number of two-level Rydberg atoms placed in a certain fixed positions in the
space in the presence of the magnetic field along $z$ axis. Then the second
order dipole-dipole perturbation will introduce the binary coupling
\[
\sum_{n<m}C\frac{\widehat{\sigma}_{z,n}\widehat{\sigma}_{z,m}}{R_{mn}^{6}},
\]
known as dipole blockade, where $R_{mn}$ stands for the distance between every
pair of atoms, and $C$ is a constant accounting for the atomic dipole
susceptibility. The higher order interactions, say the three-body and the
four-body ones, in principle can also be induced, though with much more
involved procedures relying on the multiphoton processes based on virtual
transitions to the neighboring Rydberg states, different from the two levels
under consideration, that would favorize the higher-order nonlinear
susceptibilities. This process may equally include addressing and manipulation
of the position of individual atoms by implementing optical tweezers. These
actions can be seen here as construction of the effective interatomic
interaction Hamiltonian $F\left(  \left\{  \widehat{\sigma}_{z,n}\right\}
\right)  $ of Eq.(\ref{Eq1}).

In an external RF electromagnetic field of strength $\mathcal{E}$ that has
typical frequency domain around the transition frequency $\omega$ of the two
level system, with the interaction Hamiltonian
\begin{equation}
\widehat{V}=\mathcal{E}d\sum_{n=1}^{N}\widehat{\sigma}_{x,n}, \label{IntHam}%
\end{equation}
quantum transition among the states of the ensemble can occur. Here $d$
denotes the transition dipole moment matrix element. Since the transition
frequency $\omega$ belongs to the RF domain, with the typical wavelength
exceeding the size of the atomic ensemble, all the atoms found themselves in
the field of the same strength, which is therefore is placed in front of the summation.

\begin{figure}
[th]
\begin{center}
\includegraphics[
height=1.3318in,
width=2.872in
]%
{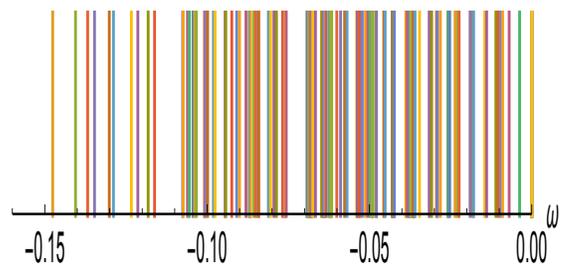}%
\caption{Absorption spectrum of multiphoton transitions calculated for 7 atoms
interacting via the second-order van der Waals coupling $1/R^{6}$ that are
placed in a unit 3D box with periodic boundary conditions. The frequency axis
(in arbitrary units) The multiphoton resonance corresponding to the minimal
energy per atom is at the red side. }%
\label{Fig1b}%
\end{center}
\end{figure}

The absorption spectrum of the ensemble with the energy spectrum of
Fig.\ref{Fig1a}\ is depicted in Fig.\ref{Fig1b}.%
On the red side of the single-atom-single-photon transition frequency, one can
find multiphoton resonances of different orders $m$, and among them one,
located at the left edge of this multiphoton absorption spectrum. This very
resonance corresponds to the transition to the state with the minimum energy
per exited atom. Our aim to put our ensemble there.

However, the \ composite matrix element
\begin{equation}
V_{m}=\sum_{\substack{\mathrm{all}\\\mathrm{excitation}\\\mathrm{channels}%
}}\frac{\left(  \mathcal{E}d\right)  ^{m}}{\prod
\limits_{\substack{\mathrm{all}\\_{{}}\mathrm{intermediate}\\_{{}%
}\mathrm{detuninings}}}\left(  k\hbar\omega-E_{s}\right)  } \label{ComME}%
\end{equation}
of the $m$-photon transition from the ground state to this multiply excited
state of the ensemble should be very small, such that the transition in
question may require a very long time. For a classical-school multiphoton
spectroscopist, the suggestion to carry out, say, a 100-photon transition in
an ensemble of 200 atoms a might sound wild, -- the Rabi frequency of such
transition scales the intensity to the power of the half of the number of
required excitations, an is an extremely small quantity. But in the Quantum
Computation Science, the exponentially long lasting processes are rather usual
things, the question is only in the minimization of the gross rate of the time
required with the number of the two-level systems involved. If the energy
position of the state to be populated is known, and there are no other state
intervening the process, the situation can be considered as favorable, the
Rabi transition period to the isolated target level scales as the square root
of the total number of the states in the system, corresponding to the
continuous analog of the Grover's search algorithm\cite{Grover}.

In a sense, performing a multiphoton transition of a very small amplitude,
which does not require anything else but keeping coherence during a very long
time, does not look as a process more difficult than keeping the coherence and
performing in the same time a complicated control of a quantum computer
required for realization of the Grover's computational algorithm. We therefore
propose, if we may, to call "the Grover's time" the Rabi period of the
transition to an isolated level of a complex spectrum of a quantum computer,
while considering it as a process of multiphoton population of the eigenstates
of time-independent Hamiltonian. The Grover's time is thus the parameter to be
compared with another time parameter of a complex spectrum -- the Heisenberg
return time, given by the state density multiplied \ by the Plank constant,
the time when one starts to distinguish discreet and continuous spectrum.

Generally speaking, the only condition which limits the selectivity of the
transition to an isolated level is the time-energy uncertainty principle, --
the transition time should not be shorter than the inverse size of the gap
separating the target level from the other levels. Or in other words, the
Grover's time has to be longer as compared to the Heisenberg time.
Unfortunately, in the general case, one can hardly rich the such a regime
where the excitation \ process conforms the model of multiphoton transitions
between just two levels. There are at least two players that may intervene the
process. First of all, there might be no sufficiently large energy gap
separating the lowest in frequency resonance from the other levels. In this
case, the population transfer may occur not to a single state, but to a large
group of the states, each of which by itself is a rather good approximation to
the required solution of the mathematical problem. However, dynamics of such a
process is distinct from that of the two-level system and might result in
broadening of the population distribution over the energy scale.

\begin{figure}
[h]
\begin{center}
\includegraphics[
height=2.8089in,
width=1.4261in
]%
{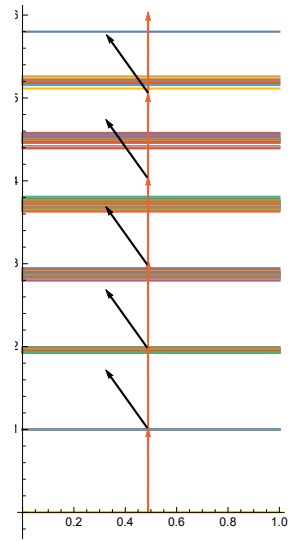}%
\caption{Raman excitation of the band edges of the atomic ensemble permits to
address the states with specified total number of the excited atoms. Virtual
transitions induced by a strong blue-detuned field are shown by red arrows.
Blue-detuned weak field induces transitions to a chosen edge of the spectrum
(black arrows). In this figure, for the case of 7 atoms we show the regime
where the number of excitations equals 3. For a larger atomic system, The
edges of the states with the total number of excitations equal multiples of 3
will also be close to the Raman resonance.}%
\label{Fig2}%
\end{center}
\end{figure}

The second reason is coexistence of the weak highly multiphoton resonant
transitions to the target levels at the absorption spectrum edge with strong
but detuned resonances corresponding to the states that are much more far from
the edge but that are much stronger coupled to the ground state since they
require less photons for the transitions. They also may spoil the width of the
energy distribution making it much larger than this would be according to the
energy-time uncertainty principle. To a certain extend, the role of such
"spoiler" transitions can be reduced by employing Raman excitation scheme, as
shown in Fig.\ref{Fig2}.%
By applying two external fields, one detuned to the blue side, and the other
detuned to the red side relative to the frequency of the single atom
transition, one can specifically address the state of the ensemble with a
chosen minimum number of the total excitations and the numbers of excitations
given by multiples of this minimum number. In such a way, certain spoiler
transitions can be moved far out from the resonance.

\section{Dynamics of a level-band system.}

We remind here the main features of the evolution of the level-band system and
present some pictures helping to qualitatively understand meaning of the
parameters responsible for different regimes. In the atomic units, the
Schr\"{o}dinger equation for a single level interacting with a band of $M$
levels reads%
\begin{align}
i\frac{\partial}{\partial t}\psi_{0}  &  =E_{0}\psi_{0}+\sum_{n=1}^{M}%
V_{n}\psi_{n},\label{SchEq}\\
i\frac{\partial}{\partial t}\psi_{n}  &  =E_{n}\psi_{n}+V_{n}\psi
_{0},\nonumber
\end{align}
where the coupling matrix elements $V_{n}$ are chosen real. For the initial
condition $\psi_{0}\left(  t=0\right)  =1$ imposed on the amplitude of the
single level, and after taking direct and inverse Fourier transforms,
Eq.(\ref{SchEq}) yields the exact solutions%
\begin{align}
\psi_{0}  &  =\frac{1}{2\pi i}\int\frac{e^{-i\varepsilon t}}{\varepsilon
-E_{0}-\sum_{m=1}^{M}\frac{V_{m}^{2}}{\varepsilon-E_{m}}}d\varepsilon
,\label{SchEqSol}\\
\psi_{n}  &  =\frac{1}{2\pi i}\int\frac{e^{-i\varepsilon t}}{\left(
\varepsilon-E_{0}-\sum_{m=1}^{M}\frac{V_{m}^{2}}{\varepsilon-E_{m}}\right)
\left(  \varepsilon-E_{n}\right)  }d\varepsilon,\nonumber
\end{align}
with the integration contour going from $\varepsilon=-\infty+io$ to
\ $\varepsilon=\infty+io$, where $o=+0$.

At times shorter than the Heisenberg return time given by the inverse of the
mean band level spacing, according the uncertainty principle, the summation
can be replaced by integration. For a band infinitely broad in both positive
and negative direction, assuming constant $V_{n}$ one obtains%
\begin{align}
\sum_{m=1}^{M}\frac{V_{m}^{2}}{\varepsilon-E_{m}}  &  \rightarrow gV^{2}%
\int\frac{dE_{m}}{\varepsilon-E_{m}}=-i\pi gV^{2}\nonumber\\
\psi_{0}  &  =e^{-\pi gV^{2}t}\label{FGR}\\
\psi_{n}  &  =\frac{e^{-\pi gV^{2}t}-e^{-iE_{n}t}}{E_{0}-E_{n}-i\pi gV^{2}%
},\nonumber
\end{align}
where $g$ is the band state energy density. In other words, the level
exponentially looses its population $\rho_{0}=\left\vert \psi_{0}\left(
t\right)  \right\vert ^{2}$ in accordance with the Fermi "golden rule" with
rate $2\pi gV^{2}$ .This population gets distributed among the resonant band
levels following the Cauchy profile%
\begin{equation}
\rho_{n}=\frac{1}{\left(  E_{0}-E_{n}\right)  ^{2}+\left(  \pi gV^{2}\right)
^{2}}. \label{EnDis}%
\end{equation}

At the times shorter than the Heisenberg time, the discreet spectrum can be
considered continuous. One can thus call it quasicontinuum. At the times
longer than the Heisenberg times, replacement of the sum by the integral is
not valid, and the exponential level decay no longer occur. On the contrary,
the population experience recurrences -- partially it returns back to the
level and becoming of the order of the population of all other resonant levels
it manifests fluctuations that depend on the specific positions of the band
levels. The level even does not decay on average, when the Heisenberg return
time is shorter than the Fermi golden rule decay time. This is the case of the
couplings smaller than the average distance among the band energy levels.

\begin{figure}
[h]
\begin{center}
\includegraphics[
height=1.7556in,
width=2.8573in
]%
{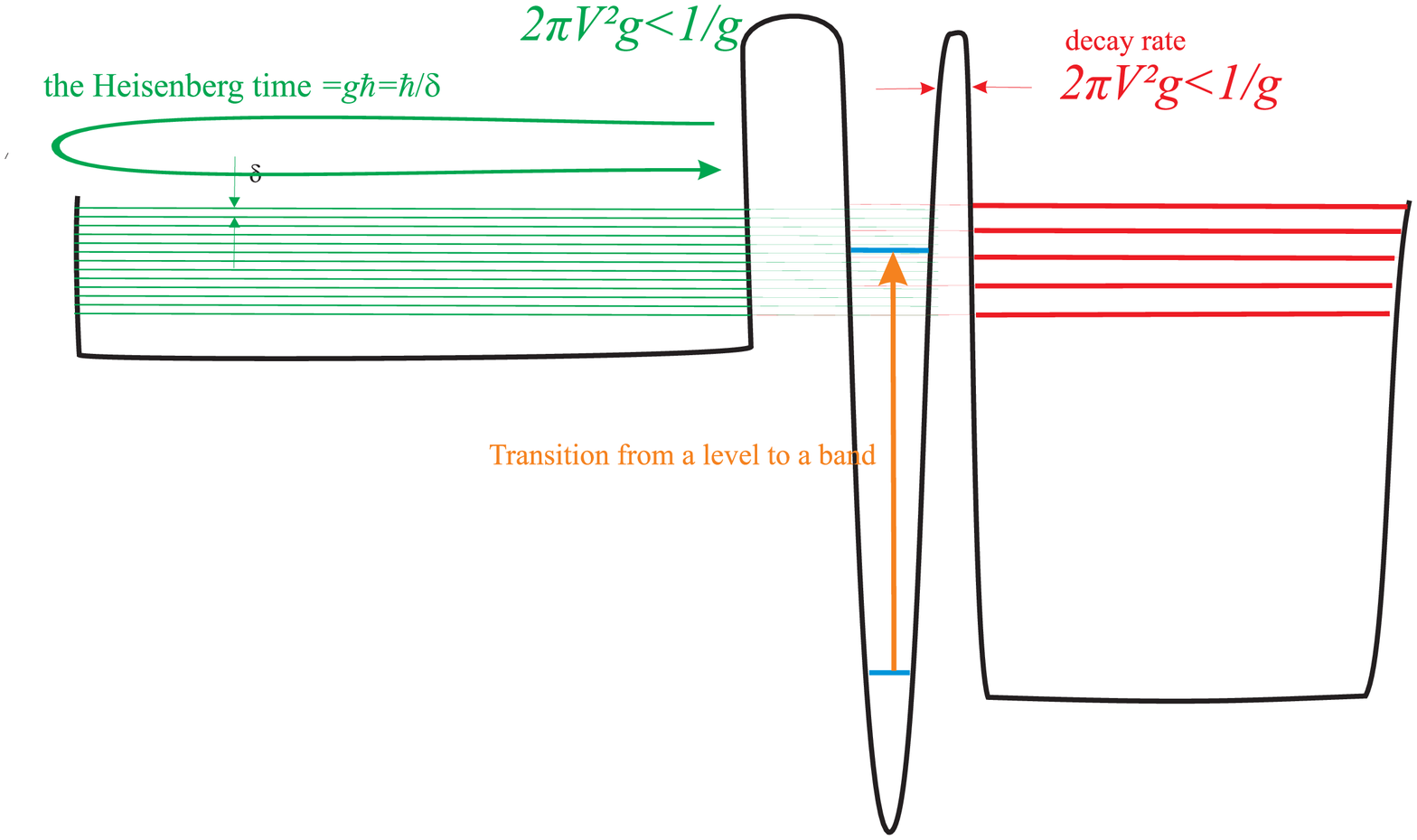}%
\caption{Transition from a level to a band can be seen as the excitation of a
quantum level in a narrow potential pit that can decay via tunneling to the
domains of the periodic motions in other , broad, potential pits. Red levels
in the narrower pit correspond to a shorter oscillation period, that is to a
shorter Heisenberg return time, whereas the green levels correspond to a slower
motion. Each pit represents the state with a given coupling size. The Fermi
golden rule decay rate is proportional to the tunnel transparency of the
barriers. If the periods of the motion in the broad pits corresponding to the
Heisenberg return times are shorter than the decay rate, the decay does not
occur. However, if there are many (not just two as in the picture) potential
pits are available for the tunneling, the decay still may ocuure.}%
\label{Fig4}%
\end{center}
\end{figure}

Now the question arises -- what happens to the essentially inhomogeneous band,
when the couplings $V$ can differ by orders of magnitudes for neighboring band
levels? The average squared interaction and the average state density are no
longer the parameters that govern the population dynamics. The reason for this
is rather clear, -- the main contribution to the mean square coupling may come
from the rare strongly coupled band levels (the spoiler states) while the main
contribution to the state density might come from the extremely weak-coupled
states. In other words, the Fermi "golden rule" transition rates and the
Heisenberg return times may differ drastically for the strongly and weakly
coupled submanifolds of the band levels. This situation is illustrated in
Fig.\ref{Fig4} by an "eye-guiding" example of two potential pits with the
spectra of the oscillatory motion, that can be accessed via tunneling from an
excited state of a narrow potential pit.%

The situation may even turn out to be such, that non of many different
submanifolds conforms the requirement of the "golden rule" applicability, and
still all the submanifolds together will be able to accommodate all the
level's population. One therefore needs to find suitable statistical
characteristics of the band levels that govern the population dynamics and the
energy distribution. One of the authors (V.A.) has considered such a problem
in the context of multiphoton vibrational laser excitation of the polyatomic
molecules\cite{Akulin}, and (in collaboration) in the context of the
excitation exchange in the ensemble of cold Rydberg atoms\cite{Mourachko}. In
seems \ expedient to briefly remind the results, since a similar situation
occurs at the quasicontinnum edge.

The band is considered as a manifold of levels randomly placed within a broad
energy strip, with independent statistics of the level energy positions
$E_{n}$ and the sizes of the coupling matrix elements $V_{n}$. Moreover,
position of each band level is assumed to be statistically independent from
the positions of others. This assumption is very much different from the usual
model of complex spectra, that conform models of Gaussian orthogonal, unitary
or simplectic ensembles. However, for the situation of a complex system
composed of the elements interacting via committing operators, which is the
case of the Hamiltonian Eq.(\ref{Eq1}) with the interaction $F\left(  \left\{
\widehat{\sigma}_{z,n}\right\}  \right)  $, the model of the statistically
independent level position uniformly distributed over an energy strip seems
much more adequate, since the perturbation by commuting operators does not
produce level "repulsion".

For the power-law statistics of the couplings $g\left(  V\right)  \sim
V^{-\alpha}$, one can identify three different regimes of the level decay. For
$\alpha<2$ the expectation value of the number of resonances, he number of the
band levels that satisfy the condition of the detuning smaller than the
coupling is finite, -- $\int V$ $g\left(  V\right)  dV<\infty$. Therefore the
number of band states accessible for the population transfer is finite, and
the level does not decay completely. The transferred part of the level
population is distributed among the levels with strong coupling. The main part
of the dense but weakly coupled band levels remains unpopulated.

For $2<\alpha<3$ the level transfers its population completely, since the
expectation number of resonances is infinite. The main contribution to this
expectation value comes from the weakly bounded states, that at the end
receive all the population of the level. However the process goes very slow,
according not to the exponential, but to the power law $t^{-\beta\left(
\alpha\right)  }$ . The population first goes to the strongly coupled levels,
from where it is retrieved back to the isolated level after the corresponding
Heisenberg time and is transferred further to a group ow weaker coupled
states. This population transfer continues toward yet weaker and weaker
coupled band states with gradually decreasing velocity. For $\alpha>3$ the
process goes to the weakly coupled states, although the energy width of the
population distribution is given by Eq.(\ref{EnDis}) in accordance with the
Fermi "golden rule".

We now are going to consider similar processes at the quasicontinuum edge,
paying attention to the energy distribution of the transferred population with
the aim to make it located at the very edge of the spectrum and to be as
narrow as it is suggested by the energy-time uncertainty principle.

\section{\label{Uniformband}Population dynamics at the edge of a uniform
quasi-continuum spectrum}

Let us consider a Hamiltonian $\hat{H}_{0}$ with eigen states $\left\vert
n\right\rangle $ that for $n>0$ has homogeneous spectrum $E_{n}$ of spectral
density $g$ located at the energy axis in the interval $\left[  0,\Gamma
\right]  $ and an additional state $\left\vert 0\right\rangle $ with the
negative eigen energy $E_{0}$. The state vector is given in terms of the state
amplitudes as
\[
\left\vert \psi(t)\right\rangle =\psi_{0}\left\vert 0\right\rangle +\sum
_{n=1}^{n=N}\psi_{n}\left\vert n\right\rangle .
\]

At times, shorter than the Heisenberg return time, we can consider the
homogeneous spectrum as continuous, that is
\begin{align}
\hat{H}_{0}  &  =\left\vert 0\right\rangle E_{0}\left\langle 0\right\vert
+\sum_{n=1}^{n=N}\left\vert n\right\rangle E_{n}\left\langle n\right\vert
\nonumber\\
&  \simeq\left\vert 0\right\rangle E_{0}\left\langle 0\right\vert
+\int\limits_{0}^{\Gamma}\left\vert E\right\rangle E\left\langle E\right\vert
g\left(  E\right)  dE. \label{ap}%
\end{align}
We assume that at time $t=0$, the state $\left\vert 0\right\rangle $ is fully
populated, that is $\psi_{0}(t=0)=1$, and the interaction $\hat{V}$ \ \ that
couples the state $\left\vert 0\right\rangle $ with the rest of the spectrum,
\begin{align}
\hat{V}  &  =\sum_{n=1}^{n=N}\left(  \left\vert 0\right\rangle V_{0n}%
\left\langle n\right\vert +\left\vert n\right\rangle V_{n0}\left\langle
0\right\vert \right) \\
&  \simeq\int\limits_{0}^{\Gamma}\left(  \left\vert 0\right\rangle
V_{0E}\left\langle E\right\vert +\left\vert E\right\rangle V_{E0}\left\langle
0\right\vert \right)  dE.
\end{align}
with $V_{n~0}=V_{0~n}^{\ast}$, is switched on abruptly at this moment. For
this case the state vector reads
\[
\left\vert \psi(t)\right\rangle =\psi_{0}\left(  t\right)  \left\vert
0\right\rangle +\int\limits_{0}^{\Gamma}\psi_{E}(t)\left\vert E\right\rangle
dE.
\]

We first consider the transfer of population and its distribution $\rho
_{E}(t)=\left\vert \psi_{E}(t)\right\vert ^{2}$ at the red-edge of the uniform
spectrum of the Hamiltonian $H_{0}$ after this switching, assuming also that
the coupling elements are real and uniform, i.e., $V_{E0}=V$. Our aim is to
understand how the presence of the edge modifies the population dynamics
compared to the case of the infinite band. To this end we write the
Schr\"{o}dinger equation for the Fourier transforms $\psi_{j}(\varepsilon
)=\int\psi_{j}(t)e^{i\varepsilon t}dt$ of the amplitudes $\psi_{j}(t)$ and
arrive at the set of algebraic equations
\begin{align}
\varepsilon\psi_{0}(\varepsilon)  &  =E_{0}\psi_{0}(\varepsilon)+\sum
_{n=1}^{N}V\psi_{n}(\varepsilon)+1\\
\varepsilon\psi_{n}(\varepsilon)  &  =E_{n}\psi_{n}(\varepsilon)+V\psi
_{0}(\varepsilon).
\end{align}
One straightforwardly finds the solutions
\begin{align}
\psi_{0}(\varepsilon)  &  =\frac{1}{\varepsilon-E_{0}+V^{2}\sum_{m=1}^{N}%
\frac{1}{E_{m}-\varepsilon}}\label{ShEqEd1}\\
\psi_{n}(\varepsilon)  &  =\frac{V}{\left(  \varepsilon-E_{n}\right)  \left(
\varepsilon-E_{0}+V^{2}\sum_{m=1}^{N}\frac{1}{E_{m}-\varepsilon}\right)  }~~.
\label{SchEqEd2}%
\end{align}

At this point, we make use of the fact that the time is short as compared to
the Heisenberg return time and replace the sums by the integral
\begin{align}
V^{2}g\int\limits_{0}^{\Gamma}\frac{1}{E-\varepsilon}dE  &  =V^{2}g\left(
\log(\Gamma-\varepsilon)-\log(-\varepsilon)\right) \\
&  \simeq V^{2}g\left(  \log\Gamma-\log(-\varepsilon)\right)
\end{align}
where $\Gamma$ has meaning of the cut-off energy limiting the size of the band
affected by the interaction $V$. We also set $\log(\Gamma-\varepsilon
)\approx\log(\Gamma)$. This assumption just results in a small deviation from
precise results at short times (corresponding to large $\varepsilon$), but it
considerably simplifies the calculations allowing one to ignore the
contribution of the branching point at $\varepsilon=\Gamma$ when performing
the inverse Fourier transformation. We denote $V^{2}g=w$ and, define the
positive quantity $\bar{E}_{0}=-E_{0}+w\log(\Gamma)$, thus arriving at
\begin{align}
\psi_{0}(\varepsilon)  &  =\frac{1}{\varepsilon-w\log(-\varepsilon)+\bar
{E}_{0}}\label{b1}\\
\psi_{E}(\varepsilon)  &  =\frac{V}{\left(  \varepsilon-E\right)  \left(
\varepsilon-w\log(-\varepsilon)+\bar{E}_{0}\right)  }, \label{b2}%
\end{align}
which after the inverse Fourier transform yields the result
\begin{align}
\psi_{0}(t)  &  =\frac{1}{2\pi}\int\limits_{-\infty+i0}^{\infty+i0}%
\frac{e^{-i\varepsilon t}}{\varepsilon-w\log(-\varepsilon)+\bar{E}_{0}%
}d\varepsilon\label{f1}\\
\psi_{E}(t)  &  =\frac{1}{2\pi}\int\limits_{-\infty+i0}^{\infty+i0}%
\frac{Ve^{-i\varepsilon t}}{\left(  \varepsilon-E\right)  \left(
\varepsilon-w\log(-\varepsilon)+\bar{E}_{0}\right)  }d\varepsilon~~.
\label{f2}%
\end{align}

\begin{figure}
[h]
\begin{center}
\includegraphics[
height=2.1837in,
width=2.9092in
]%
{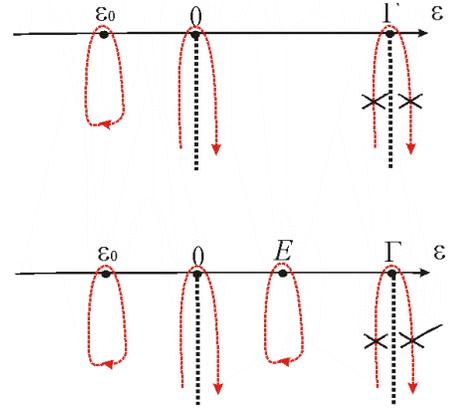}%
\caption{Integration contours in the complex plane of $\varepsilon$ (red
curves) of the inverse Fourier transformation for $\psi_{0}$ (upper contour)
and for $\psi_{E}$ (lower contour). Dotted lines show cuts of the Riman
surfaces starting \ at the logarithmic branching points; Contribution of the
branching point at the upper edge of the band is ignored at times
$t>1/\Upsilon$.}%
\label{Cont}%
\end{center}
\end{figure}

Now we turn to the calculation of the integrals of the inverse Fourier
transformation. To assign the contour to the lower-half of the complex plane,
we first set the branch-cut of the $\log$ function along the negative
imaginary axis. The integration contour is shown in Fig.\ref{Cont}.%
The pole for the integral in Eq.(\ref{f1}), is found to be the negative real
number $\varepsilon_{0}=-wW_{0}\left(  \frac{e^{\bar{E}_{0}/w}}{w}\right)  $
where $W_{0}(z)$ denotes the principal branch of the Lambert function, also
known as product logarithmic function. The point $\varepsilon_{0}$ is the root
of the integrand denominator. Residue calculus and the allowance for the
contributions of the integrals along the branch cut result in
\begin{equation}
\psi_{0}(t)=-i\frac{e^{-i\varepsilon_{0}t}}{1-\frac{w}{\varepsilon_{0}}}-I~,
\label{g1}%
\end{equation}
where%
\begin{align}
I  &  =\frac{i}{2\pi}\int_{0}^{\infty}\frac{e^{-yt}}{iy+w\log(y)+w\frac{i\pi
}{2}-\bar{E}_{0}}dy\nonumber\\
&  -\frac{i}{2\pi}\int_{0}^{\infty}\frac{e^{-yt}}{iy+w\log(y)-w\frac{i3\pi}%
{2}-\bar{E}_{0}}dy~~. \label{g2}%
\end{align}
By analogy, for the states in the quasi-continuum, Eq.(\ref{f2}), we obtain:
\begin{align}
\psi_{E}(t)  &  =-\frac{iVe^{-i\varepsilon_{0}t}}{\left(  \varepsilon
_{0}-E\right)  (1-\frac{w}{\varepsilon_{0}})}-\nonumber\\
&  \frac{iVe^{-iEt}}{\left(  E-w\log(-E)+\bar{E}_{0}\right)  }-J, \label{h1}%
\end{align}
where
\begin{align}
&  J=\frac{iV}{2\pi}\left(  \int_{0}^{\infty}\frac{e^{-yt}}{\left(
iy+E\right)  \left(  iy+w\log y+w\frac{i\pi}{2}-\bar{E}_{0}\right)  }dy\right.
\nonumber\\
&  -\left.  \int_{0}^{\infty}\frac{e^{-yt}}{\left(  iy+E\right)  \left(
iy+w\log y-w\frac{i3\pi}{2}-\bar{E}_{0}\right)  }dy\right)  \label{h2}%
\end{align}

\begin{figure}[h]
\begin{center}
\includegraphics[
height=3.32in,
width=2.4517in
]{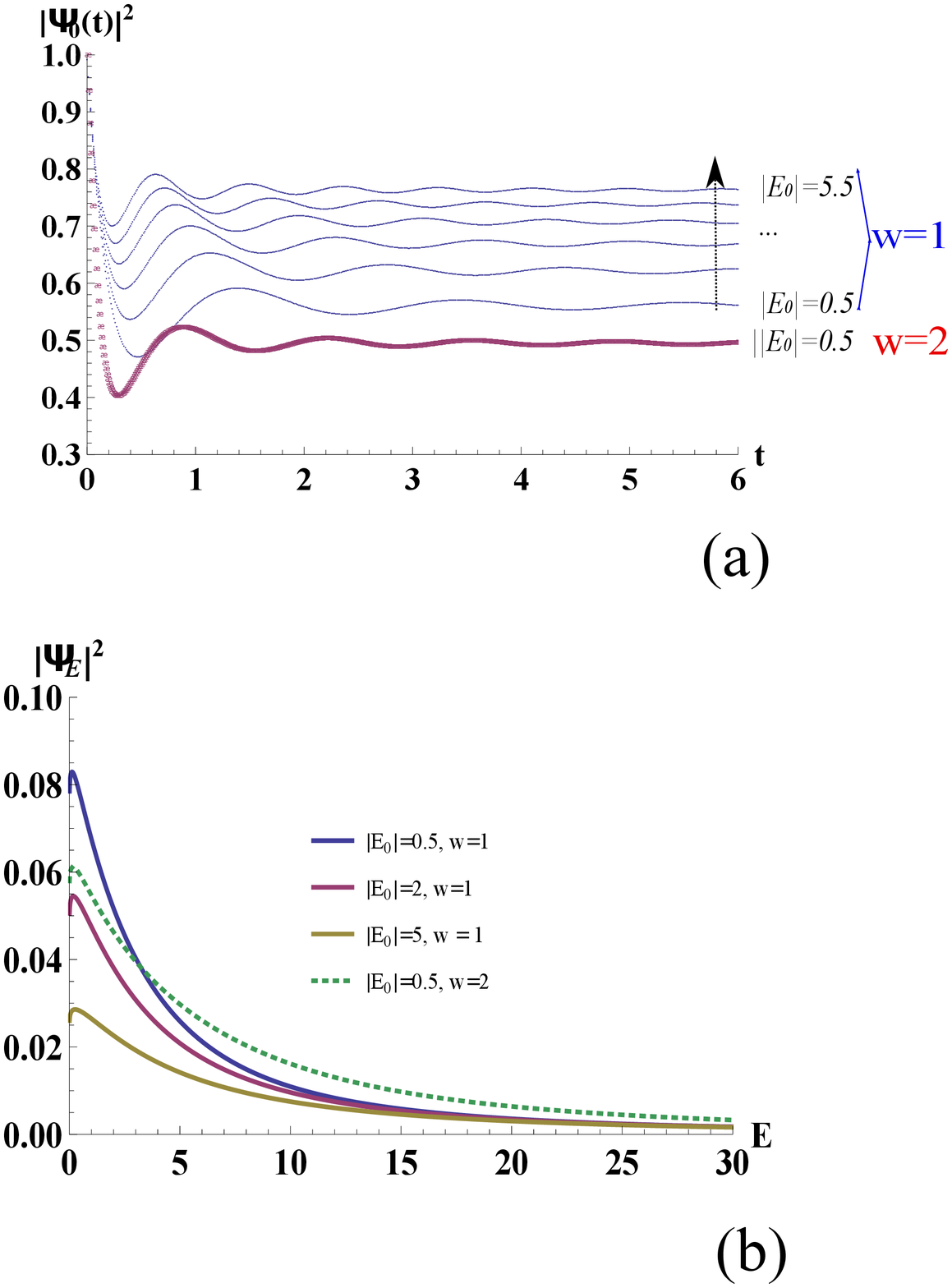}
\end{center}
\caption{\textit{(a)} The decay of the population of the isolated level as a
function of time. An exponentially fast decay rate is followed up by time
domain of slower rate as described in Eq.(\ref{NonExp}). Asymptotically the
population transferred to the continuum depends on the size of the gap
$|E_{0}|$ and $w=gV^{2}$. \textit{(b)} The distribution of the population over
the energies of the continuum for sufficiently long times, $(wt\log
wt)^{2}>>1$. For the plot we have averaged out the fast oscillating
interference/product term which results by taking the square magnitude of
Eq.(\ref{h1}). We have employed the cut-off spectrum parameter $\Gamma=40$.}%
\label{fig1}%
\end{figure}

Contribution of the poles give the asymptotic population for long times, while
the contributions resulting from the integrals $I$ and $J$ along the cuts
decrease with the time elapse. To overview the global time-dependence, in
Fig.~\ref{fig1}~(a) we present the time-evolution of isolated level population
$\left\vert \psi_{0}(t)\right\vert ^{2}$ for different parameters $w$ and
$|E_{0}|$, and where the integral $I$ of Eq.(\ref{g2}) is performed
numerically. The time dependence possess initially a fast stage with
exponential decay (as for the case where the edges are absent) that is
followed up by a regime of slower decay. To analytically identify the time
dependence in the `slow' regime we perform a change of variable, $u=yt$ in the
integrals $I$ and $J$ in Eqs. (\ref{g2}) and (\ref{h2}) respectively. One then
can see that the time dependence of the populations scales for long $t$ as
\begin{equation}
\left\vert \psi_{E}(t)\right\vert ^{2}-\left\vert \psi_{E}(\infty)\right\vert
^{2}\sim(wt\ \log\left(  wt\right)  )^{-1}. \label{NonExp}%
\end{equation}

Let us consider now the distribution of the transferred population in the
regime of sufficiently long times, $(wt\log wt)^{2}>>1$ where the contribution
of the integrals $I$ and $J$ can be safely ignored. According to
Eq.(\ref{h1}), the distribution of excited population over large energies
$E$,
\begin{align}
\rho_{n}(t)  &  =\frac{V^{2}}{\left(  \varepsilon_{0}-E_{n}\right)
^{2}(1-\frac{w}{\varepsilon_{0}})^{2}}+\label{PopUnif}\\
&  \frac{V^{2}}{\left(  E_{n}-w\log(-E_{n})+\bar{E}_{0}\right)  ^{2}%
},\nonumber
\end{align}
scales as $1/E_{n}^{2}$ for large energies $E_{n}$. Note, that the population
distribution profile is a superposition of two profiles with the inverse
square dependence on the energy. Both of them correspond to the maxima at the
negative energies, that is beyond the band, below the edge position. One of
the profiles is centered at the position $\varepsilon_{0}$, relatively close to
the band edge, while the other is centered close to the position of the
isolated level energy $\bar{E}_{0}$. The first profile, though it is scaled by
the factor $(1-\frac{w}{\varepsilon_{0}})^{-2}$, still may give a more
significant contribution to the population at the quasicontinuum edge as
compared to that of the second one. In Fig.~\ref{fig1}~(b) distributions of
population $\left\vert \psi_{E}\right\vert ^{2}$ are plotted for different
parameters and we observe that there is a clear maximum near the edge.
According to numerical studies the width of the distribution follows an
approximate dependence as $\sim w/\left\vert \overline{E}_{0}\right\vert
=gV^{2}/\left\vert \overline{E}_{0}\right\vert $.%
\begin{figure}
[h]
\begin{center}
\includegraphics[
height=2.3575in,
width=3.5362in
]%
{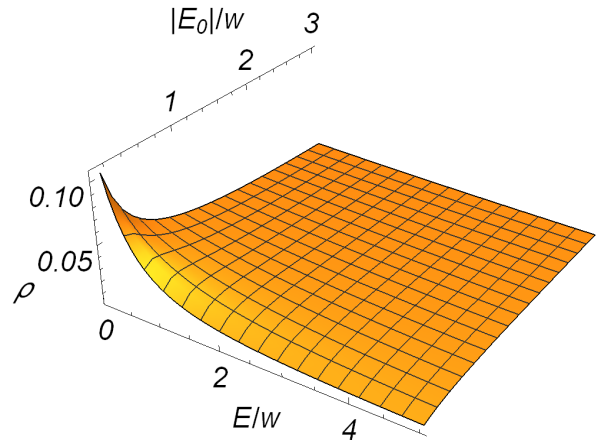}%
\caption{Thethe density of the transferee population part $\int_{0}%
^{\Gamma/100}\left\vert \psi_{E}\right\vert ^{2}dE$ as a function of the gap
$|E_{0}|/w$ and the spectral position $|E|/w$. of the band levels next to the
edge.}%
\label{fig1Bbiss}%
\end{center}
\end{figure}

Let us summarize the results of the model of uniform quasicontinuum spectrum
with edge and the abrupt switch-on of the interaction. After the switch-on,
the population initially localized at the isolated level gets partially
transferred to the quantum states at the continuum edge. Apart from the fast
($\sim e^{-wt}$) stage, the population transfer process has a slow and long
lasting stage, where the time dependence of the population transfer turns out
to be proportional to $\left(  wt\log wt\right)  ^{-1}$. The total population
transfer to the continuum is increased with increasing $w$ and decreased with
increasing gap $|E_{0}|$. The transferred population gets distributed over the
states of the quasicontinuum according to the inverse square dependence with
typical width $\sim w=gV^{2}$ corresponding to the time-energy uncertainty
principle dictated by the fast stage. In Fig.\ref{fig1Bbiss} one may overview
the overall effect of $w$ on the population transferred to the lowest part of
the spectrum together with the effect of the gap. The uniform model in this
simple setting, obviously does not provide much options for amplifying the
population near the down edge. One should just bring the isolated level as
near as possible to the edge ($\left\vert E_{0}\right\vert <<1$), choose weak
couplings $V$, and wait for long enough time such as $(wt\log wt)^{2}>>1$.

\section{\label{NonAd}Non-adiabatic transfer of the population.}

The distribution width in the problem just considered is affected by the
abrupt switch on of the interaction. The abrupt change of the interaction has
the Fourier decomposition scaling as $1/\varepsilon$, and this is the very
reason why the population distribution over energies behave as $1/E^{2}$. One
therefore can address the question : "Is it possible to make this distribution
narrower by avoiding abrupt switches?" At the first glance, one can think of
the adiabatic increase of the interaction. However, in the regime of the
highly multiphoton transitions, the adiabatic change is difficult to achieve,
since any small and slow change of the electromagnetic field strength results
in a much larger and faster change of the composite matrix element of the
multiphoton transition Eq.(\ref{ComME}). In contrast, by slow variation of the
frequency of the electromagnetic field of a constant strength one may achieve
a much more accurate control.

In order to understand the main features of the dynamics of the level-band
system subject to the interaction with slowly changing frequency, we consider
now the level-continuum edge problem in the rotating-wave approximation
assuming the parabolic time-dependence of the level position. In other words
$E_{0}(t)=E_{o}-\alpha t^{2}$, and whence the Schr\"{o}dinger equation
(\ref{SchEq}) for such a system takes the form%

\begin{align*}
i\overset{\cdot}{\psi}_{0}  &  =\left(  E_{o}-\alpha t^{2}\right)  \psi
_{0}+V\sum_{n=1}^{\infty}\psi_{n},\\
i\overset{\cdot}{\psi}_{n}  &  =E_{n}\psi_{n}+V\psi_{0}.
\end{align*}
One can implement the Laplace contour integral method that suggest the
time-frequency variable replacements%
\begin{align*}
i\frac{\partial}{\partial t}  &  \rightarrow\varepsilon,\\
t  &  \rightarrow-i\frac{\partial}{\partial\varepsilon},
\end{align*}
and obtain the Schr\"{o}dinger equation of the form%

\begin{align*}
\varepsilon\psi_{0}  &  =E_{o}\psi_{0}+\alpha\frac{\partial^{2}\psi_{0}%
}{\partial\varepsilon^{2}}+V\sum_{n=1}^{\infty}\psi_{n}\\
\varepsilon\psi_{n}  &  =E_{n}\psi_{n}+V\psi_{0},
\end{align*}
which yields the equation%

\begin{equation}
\alpha\frac{\partial^{2}\psi_{0}}{\partial\varepsilon^{2}}=\left(
\varepsilon-\sum_{n=1}^{\infty}\frac{V^{2}}{\varepsilon-E_{n}}-E_{o}\right)
\psi_{0} \label{EqScat}%
\end{equation}
for the isolated level.

In Eq.(\ref{EqScat}) one recognizes the Schr\"{o}dinger equation of a
one-dimensional quantum particle moving along the "coordinate" $\varepsilon$
in the "external potential" $\varepsilon-\sum_{n=1}^{\infty}\frac{V^{2}%
}{\varepsilon-E_{n}}$ with the total "energy" $E_{o}$, and therefore at times
shorter then the Heisenberg time, the problem of the quasi-continuum edge
excitation in the conditions of a parabolic time dependence of the detuning
can be mapped to the problem of the scattering by the potential
\begin{align*}
\varepsilon-\sum_{n=1}^{\infty}\frac{V^{2}}{\varepsilon-E_{n}}  &
\rightarrow\varepsilon+w\int\limits_{0}^{\Gamma}\frac{1}{E-\varepsilon}dE\\
&  \rightarrow\varepsilon+w\log\left(  -\varepsilon\right)
\end{align*}
where the last replacement also imply the energy scale shift $E_{o}\rightarrow
E_{o}+w\log(\Gamma)$, same as earlier.

The Schr\"{o}dinger \ equation also contains a contribution resulting from the
level motion, -- it has the form of "kinetic energy" with the parameter
$\alpha$ characterizing the rate of the parabolic approach to the minimum
detuning $E_{o}$ playing role of the inverse mass of the scattered particle.
Still, there is one important difference between such a "scattering" problem
and the regular scattering process -- the potential has an imaginary part,
which allows for the possibility of absorption of the scattering particle for
the positive coordinates.

\begin{figure}
[h]
\begin{center}
\includegraphics[
height=2.1819in,
width=3.3512in
]%
{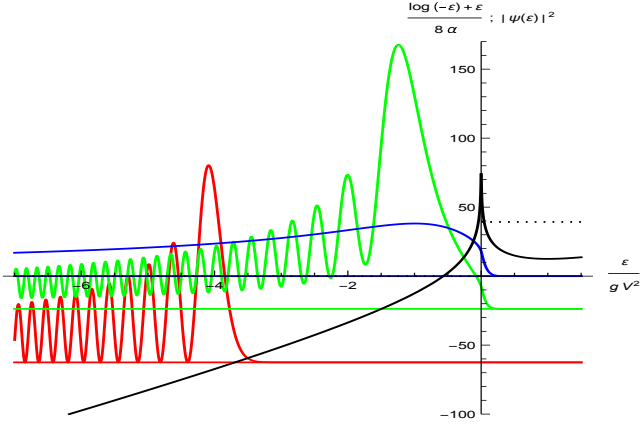}%
\caption{Non-adiabatic transition to the continuum edge as a scattering
problem in the presence of absorption. The scattering potential (solid black
curve) and the absorption profile (dotted line) result in the probability
profile depending on the "energy"of the scattered particle, which corresponds
to the minimum size $E_{o}$ of the gap between the level and the continuum
edge. For big gaps, ( negative "energies") the "absorption" , that the
probability transition to the continuum edge is negligible, and the
distribution is given by the Airy function (red curve). For small gaps the
absorption is important and the incident and back-scattered waves have
different amplitudes (green curve). For close approach, the nonadiabatic
transfer is complete (blue curve) and no "back scattered" amplitude is longer
seen. The closer sets one the "classical turning point" to the continuum edge,
the broader is the energy distribution of the absorbed population. By an
optimum choice of the "mass" $\alpha^{-1}$ , the distribution can be minimized
to the limit suggested by the uncertainty principle.}%
\label{NonAdPo}%
\end{center}
\end{figure}

We solve Eq.(\ref{EqScat}) numerically and calculate the probability profile
$\left\vert \psi_{0}\left(  \varepsilon\right)  \right\vert ^{2}$ for specific
values of the parameters $E_{o}/w$ and $\alpha/w^{2}$ . In Fig.\ref{NonAdPo}
we present the results of the calculation.%
As one can see, that the oscillations observed for negative values of the
"coordinate" $\varepsilon$, fade away for the positive coordinates. The
profile resembles Airy function, and indeed, for large detunings, that is
large negative parameters $E_{o}/w$, the profile is given by the absolute
value of the square of the Airy function, which is simply the Fourier
decomposition of a signal of constant amplitude with the parabolic frequency
modulation. No absorption occur in this limit.
\begin{figure}
[h]
\begin{center}
\includegraphics[
height=2.1958in,
width=3.2586in
]%
{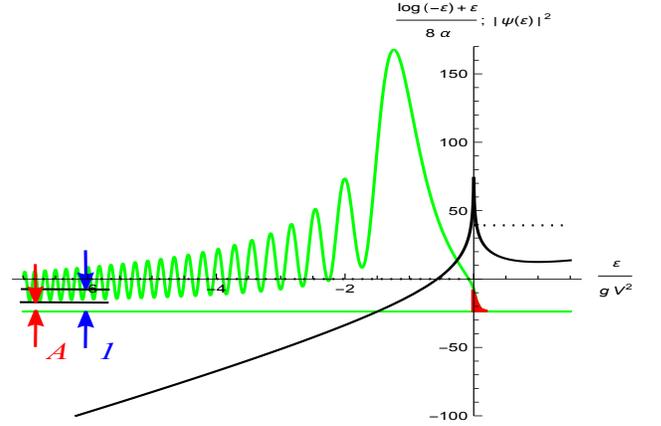}%
\caption{"Back scattered amplitude" differs from the incident amplitude due to
the absorption. The difference $A$ allows one to find numerically the amount
of the population, which has beet transferred non-adiabatically from the moving
isolated level to the band. This part corresponds to the red filled profile.}%
\label{FigD2}%
\end{center}
\end{figure}

\begin{figure}
[h]
\begin{center}
\includegraphics[
height=2.0003in,
width=3in
]%
{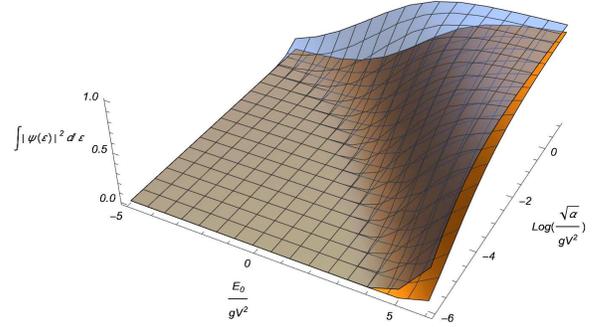}%
\caption{Probability of the nonadiabatic transition to the states at the
continuum edge calculated with WKB method (blue surface) and calculated
numerically ( yellow surface) as functions of the minimum energy gap parameter
$E_{o}/gV^{2}$ and the approach rate parameter $\alpha/\left(  gV^{2}\right)
^{2}$.}%
\label{ProbNA}%
\end{center}
\end{figure}

For decreasing minimum detuning, the exponential "tail" of the population
profile corresponding to the classically forbidden domains, starts to reach
the positive values of the coordinate, where the population can be "absorbed",
that is transferred to the edge states of the continuum. The "incident" and
the "reflected" waves no longer have equal amplitudes, and the profile
$\left\vert \psi_{0}\left(  \varepsilon\right)  \right\vert ^{2}$ no longer
reaches zero values for the negative coordinate, as it is shown in
Fig.~\ref{FigD2}. Some part $A$ of the "incident amplitude" is absorbed. The
total fraction of the population transferred to the band amounts to this very
value. In Fig.\ref{ProbNA} we show the nonadiabatic
transition probability found with the help of the WKB method and numerically,
respectively, as a function of the parameters $E_{o}/w$ and $\alpha/w^{2}$ by
solving the Schr\"{o}dinger equation (\ref{EqScat}) for scattering. For
comparison we also show the same quantity found with the help of the
semiclassical formula
\begin{equation}
\exp\left\{  -\int\limits_{W_{0}\left(  e^{-E_{o}/w}\right)  }^{0}\sqrt
{\frac{\varepsilon+V^{2}g\log\left(  -\varepsilon\right)  -E_{o}}{\alpha}%
}d\varepsilon\right\}  ,\label{SemiClassic}%
\end{equation}
with the lower integration limit given by the Lambert function. One sees, that
in the case the level slowly approaches the continuum edge, the nonadiabatic
transition probability can be rather high if $W_{0}\left(  e^{-E/w}\right)  $
is close enough to zero.%

\begin{figure}
[h]
\begin{center}
\includegraphics[
height=2.6212in,
width=3.9323in
]%
{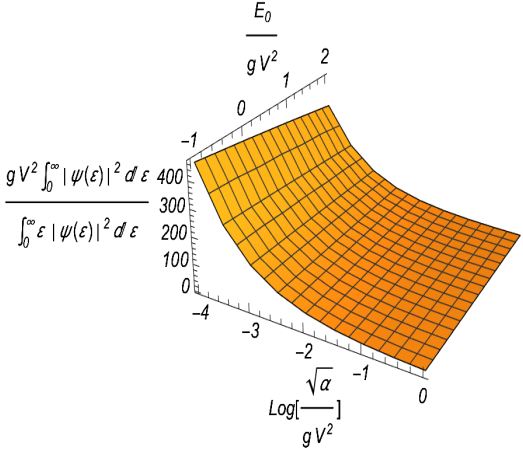}%
\caption{The inverse width of the probability energy distribution near the
continuum edge as function of the minimum energy gap parameter $E_{o}/gV^{2}$
and the approach rate parameter $\alpha/\left(  gV^{2}\right)  ^{2}$.}%
\label{PrNAWidth}%
\end{center}
\end{figure}

At the same time, the width of the population distribution may remain narrow,
since the integrand in Eq.(\ref{SemiClassic}) remains large for positive
$\varepsilon$. In Fig.\ref{PrNAWidth} we show the inverse width of the
transferred probability distribution found numerically \ as a function of the
same parameters $E_{o}/w$ and $\alpha/w^{2}$.%

One sees, that by a proper choice of the size of these parameters it is
possible to obtain a narrow energy distribution of the continuum edge, while
the probability of the non-adiabatic transition remains of the order of unit.
The distribution width, and the typical time $\sim1/\sqrt{\hbar\alpha}$
required for such a process are related by the saturation limit of the
time-energy uncertainty principle. Actually, the optimum regime corresponds to
a very slow (small $\alpha$) approach to the maximum energy $E_{o}$, which
takes a positive value. Specific values of $E_{o}$ and $\alpha$ can be
numerically found for each predetermined probability of transition\ and each
given size of the required width.

\section{Population of the quasicontinuum edge in the presence of a single
spoiler state}

The assumption of homogeneously distributed couplings over the quasicontinuum
does not look as a satisfactory model of the real spectra of multiphoton
transitions in ensembles of interacting atoms, -- neighboring states may
correspond to different orders of the multiphoton resonances and their
composite matrix elements may be drastically different. In fact, the spectral
density of the low-order resonances with high couplings has to be much smaller
as compared to the that for the much more abandoned and weakly coupled
high-order resonances.

In this Section, we make a first simple modification of the homogeneous
spectrum model. We assume the existence of a `spoiler' state $\left\vert
S\right\rangle $ of energy $E_{s}$, with $0<E_{s}<\Gamma$, whose coupling to
the ground state is much stronger than the average coupling, i.e.,
$|V_{s~0}|=|V_{s}|>>|V|$. We separately study the amplitude of the spoiler
state $\psi_{S}(t)$ and \ modify Eqs.(\ref{f1})-(\ref{f2}) as:
\begin{align}
\psi_{0}(t)  &  =\int\limits_{-\infty+i0}^{\infty+i0}\frac{d\varepsilon}{2\pi
}\frac{e^{-i\varepsilon t}}{\varepsilon-w\log(-\varepsilon)+\bar{E}_{0}%
+\frac{|V_{s}|^{2}}{E_{s}-\varepsilon}}\label{k1}\\
\psi_{E}(t)  &  =\int\limits_{-\infty+i0}^{\infty+i0}\frac{d\varepsilon}{2\pi
}\frac{Ve^{-i\varepsilon t}}{\left(  \varepsilon-E\right)  \left(
\varepsilon-w\log(-\varepsilon)+\bar{E}_{0}+\frac{|V_{s}|^{2}}{E_{s}%
-\varepsilon}\right)  }\label{k2}\\
\psi_{E_{s}}(t)  &  =\int\limits_{-\infty+i0}^{\infty+i0}\frac{d\varepsilon
}{2\pi}\frac{V_{s}e^{-i\varepsilon t}}{\left(  \varepsilon-E_{s}\right)
\left(  \varepsilon-w\log(-\varepsilon)+\bar{E}_{0}\right)  -|V_{s}|^{2}}
\label{k3}%
\end{align}
where it is implicit that $\psi_{E}=\psi_{E\neq E_{s}}$.

\begin{figure}
[h]
\begin{center}
\includegraphics[
height=2.0058in,
width=2.6725in
]%
{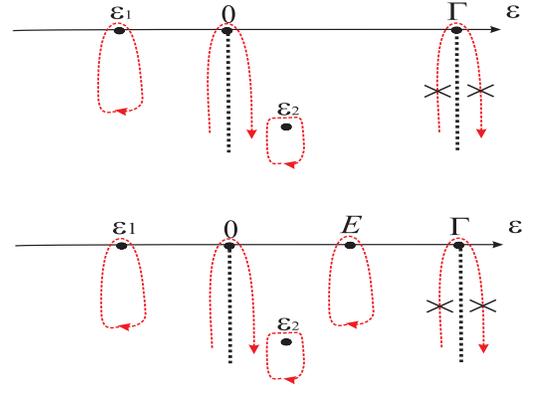}%
\caption{Integration contours for the level (upper) and the band state (lower)
amplitudes in the presence of a spoiler state with the energy $E_{s}$.}%
\label{Cont2}%
\end{center}
\end{figure}

As earlier, in order to calculate the integrals, we perform contour
integration at the complex plane. We start with the ground state amplitude
$\psi_{0}(t)$, for which the integrand denominator has two roots to be found
numerically: a real root $\varepsilon_{1}<0$ and a complex root $\varepsilon
_{2}$ which has positive real part and (small) negative imaginary part. The
integration contour as shown in Fig.\ref{Cont2}%
We integrate on the complex plane along a path that is slightly upper than the
real axis and that goes around the branch cut of the log-function -- which is
placed as before along the negative imaginary axis. Thus both residues,
corresponding to the roots $\varepsilon_{1,2}$ need to be taken into account,
arriving at:
\begin{equation}
\psi_{0}(t)=-i\sum_{k=1}^{2}\frac{e^{-i\varepsilon_{k}t}}{1-\frac
{w}{\varepsilon_{k}}+\frac{|V_{s}|^{2}}{\left(  E_{s}-\varepsilon_{k}\right)
^{2}}}-L,
\end{equation}
where
\begin{align}
&  L=\int_{0}^{\infty}\left(  \frac{ie^{-yt}}{iy+w\log(y)+\frac{i\pi w}%
{2}-\bar{E}_{0}-\frac{|V_{s}|^{2}}{E_{s}+iy}}\right. \\
&  -\left.  \frac{ie^{-yt}}{iy+w\log(y)-\frac{i3\pi w}{2}-\bar{E}_{0}%
-\frac{|V_{s}|^{2}}{E_{s}+iy}}\right)  \frac{dy}{2\pi}~.
\end{align}

\begin{figure}
[h]
\begin{center}
\includegraphics[
height=4.2281in,
width=3.1998in
]%
{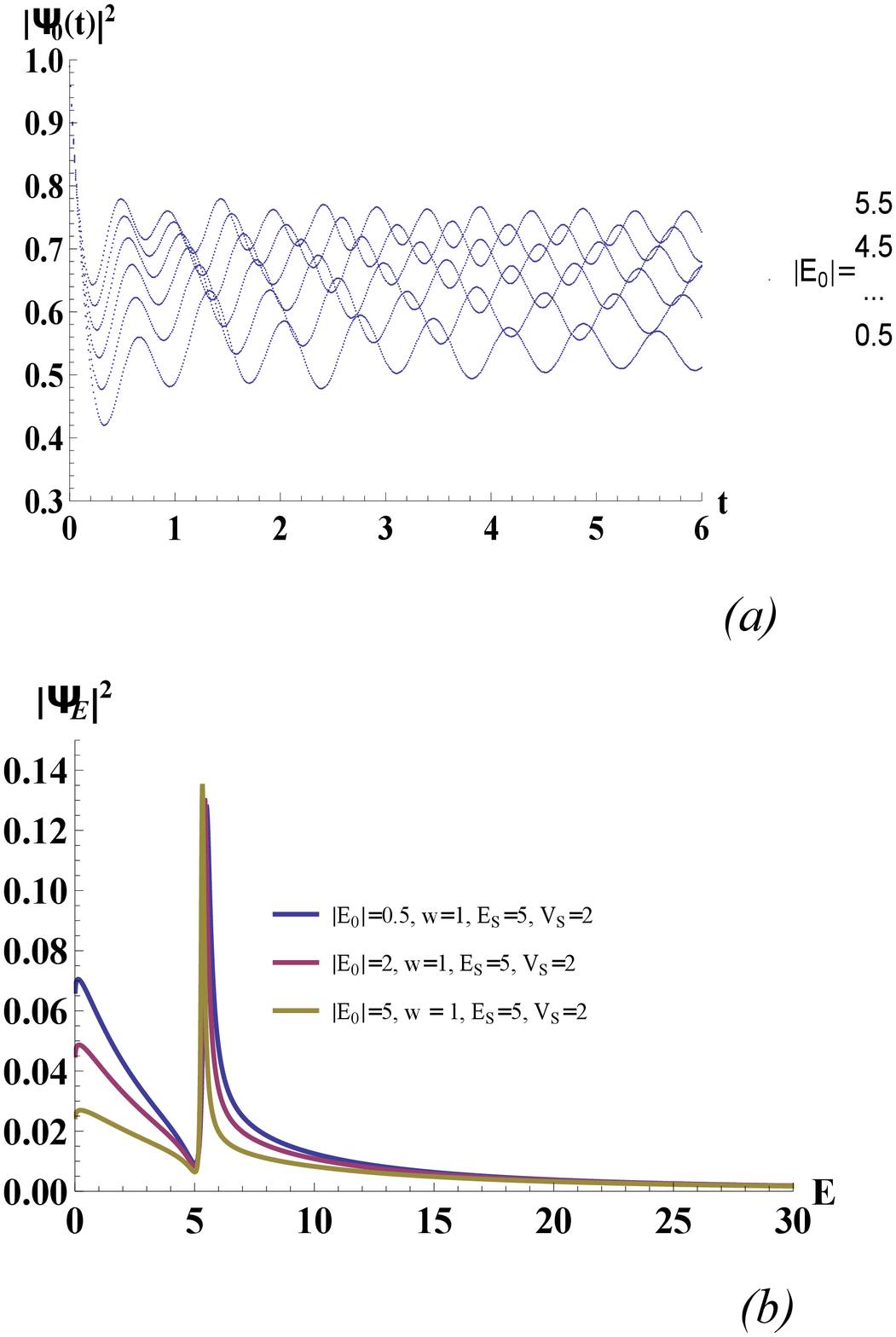}%
\caption{hh}%
\label{fig2}%
\end{center}
\end{figure}

The two regimes of the time-evolution of the population of the ground state
persist in the presence of the spoiler state, see Fig~\ref{fig2}~(a), while
there is an additional weak exponential decay due to $\varepsilon_{2}$ that is
not evident in this figure. We proceed with deriving the population
transferred to the continuum, excluding the part `absorbed' by the spoiler
state. Here, we include the additional to $\varepsilon_{1,2}$, positive pole
$E$ and arrive at:%
\begin{align}
\psi_{E}(t)  &  =\frac{-i V^{\ast}e^{-i E t}}{\left(  E-w\log(-E)+\bar{E}%
_{0}+\frac{|v_{s}|^{2}}{E_{s}-E}\right)  }\\
&  -\sum_{k=1}^{2}\frac{i V^{\ast}e^{-i\varepsilon_{k}t}}{\left(
\varepsilon_{k}-E\right)  \left(  1-\frac{w}{\varepsilon_{k}}+\frac
{|V_{s}|^{2}}{\left(  E_{s}-\varepsilon_{k}\right)  ^{2}}\right)  }-M~~,
\end{align}
where
\begin{align*}
M  &  =\int_{0}^{\infty}\frac{e^{-yt}}{iy+E}\left(  \frac{i V^{\ast}}{iy+w\log
y-\frac{i3\pi w}{2}-\bar{E}_{0}-\frac{|V_{s}|^{2}}{E_{s}+iy}}\right. \\
&  -\left.  \frac{i V^{\ast}}{iy+w\log y+\frac{i\pi w}{2}-\bar{E}_{0}%
-\frac{|v_{s}|^{2}}{E_{s}+iy}}\right)  \frac{dy}{2\pi}~.
\end{align*}

Finally we provide the population of the spoiler state where we have only the
contribution of the two poles $\varepsilon_{1,2}$,
\begin{equation}
\psi_{E_{s}}(t)=\sum_{k=1}^{2}\frac{-i V_{s}^{\ast}e^{-i\varepsilon_{k}t}%
}{\left(  \varepsilon_{k}-E\right)  \left(  1-\frac{w}{\varepsilon_{k}%
}\right)  +\varepsilon_{k}-\log(-\varepsilon_{k})+\bar{E}_{0}}-P,
\end{equation}
where
\begin{align}
~P  &  =\int_{0}^{\infty}\left(  \frac{i V_{s}^{\ast}e^{-yt}}{(iy+E_{s}%
)(iy+w\log y-\frac{i3\pi w}{2}-\bar{E}_{0})-|V_{s}|^{2}}\right. \\
&  -\left.  \frac{i V_{s}^{\ast}e^{-yt}}{(iy+E_{s})(iy+w\log y+\frac{i\pi
w}{2}-\bar{E}_{0})-|V_{s}|^{2}}\right)  \frac{dy}{2\pi}~~.
\end{align}

From the plots in Fig.~\ref{fig2}~(b) (and by comparison to those of
Fig.~\ref{fig1}) we may observe that the presence of the spoiler state does
not have a considerable impact on the shape of the distribution at the
``bottom'' of the continuum where we are interested in. Still a more detailed
numerical study shows that there is an increase of the population at the
bottom-edge as the energy $E_{S}$ spoiler state is approaching the down edge.
This is explained by the Fano-type profile in the distribution, see
Fig.~\ref{fig2}~(b), where the band states near the spoiler state get
populated and the population has a ``tail'' directed to the side of higher
energies. We note here that the `Fano' profile appearing in our case study is
due to the \textit{indirect} interaction of the spoiler state with the band
levels via the isolated level, and not due to the direct interaction as in the
original problem investigated by Fano.

\section{Spoiler state in the case of the nonadiabatic population transfer.}

\begin{figure}
[h]
\begin{center}
\includegraphics[
height=1.9303in,
width=2.898in
]%
{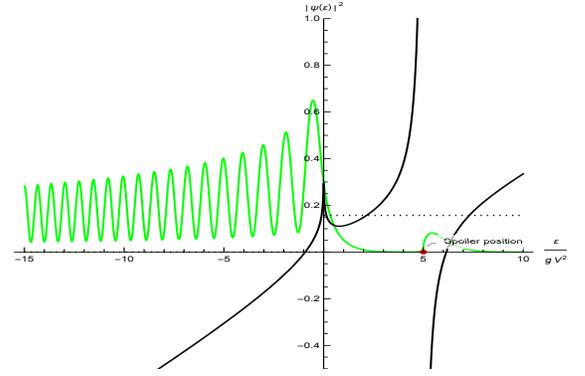}%
\caption{Numerical solution of the Schr\"{o}dinger equation (green curve)
yields the population distribution, that apart of the populated domain near
the quasicontinuum edge, contains the part located in the vicinity of the
spoiler state $E_{s}=5w$ , $V_{s}=w\sqrt{5}$ The spoiler modifies the
"scattering potential" (solid black curve for the real part and dotted black
line for the imaginary part) and creates a "classically allowed" domain at
the blue side of it's energy position.}%
\label{FigD3}%
\end{center}
\end{figure}

Now the effect of the spoiler state for the case of slowly moving level. The
Schr\"{o}dinger equation takes the form:%

\[
\alpha\frac{\partial^{2}\psi_{0}\left(  \varepsilon\right)  }{\partial
\varepsilon^{2}}=\left(  \varepsilon-E_{o}-\int\frac{1}{\varepsilon-E_{n}%
}dE_{n}-\frac{|v_{s}|^{2}}{\varepsilon-E_{s}}\right)  \psi_{0}\left(
\varepsilon\right)  ,
\]
and the new scattering potential now includes two more parameters -- the
energy position $E_{s}$ of the spoiler state, and it's "force" $|v_{s}|^{2}$.
The typical shape of the population distribution $\left\vert \psi_{0}\left(
\varepsilon\right)  \right\vert ^{2}$ following from the numerical solution of
this equation has the form depicted in Fig.\ref{FigD3}.%
As one sees, a fairly strong spoiler significantly changes the population
distribution profile. The main effect is due to the change of the "scattering
potential", -- next to the spoiler, it may result in a "classically allowed"
domain, which takes an appreciable fraction of the population.

\begin{figure}
[h]
\begin{center}
\includegraphics[
height=2.0003in,
width=3in
]%
{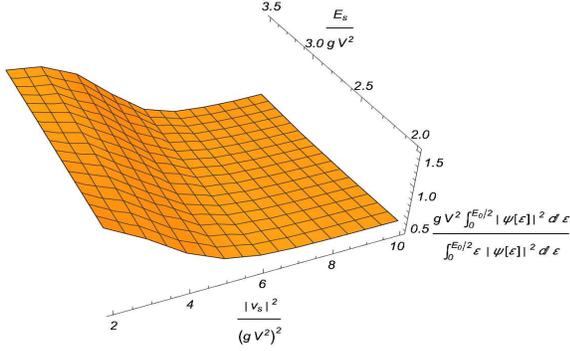}%
\caption{Dependence of the population distribution inverse width at the the
edge as function of the parameters $E_{s}$ and $|v_{s}|^{2}$ for $\left\vert
E_{0}\right\vert =5gV^{2}$ and $\alpha=\left(  gV^{2}\right)  ^{2}$ .}%
\label{FigSpNonad}%
\end{center}
\end{figure}

Two questions should be answered now: (i) How does the spoiler presence affect
the width of the distribution near the continuum edge? and (ii) How large is
the fraction of the population that has been moved in the vicinity of the
spoiler state? We start with the first question and find numerically the
dependence of the inverse of the edge population width as a function of the
parameters $E_{s}$ and $|v_{s}|^{2}.$ In Fig.\ref{FigSpNonad} we depict the
results found for the spoiler parameters $\left\vert E_{0}\right\vert
=9gV^{2}$ and $\alpha=10^{-2}\left(  gV^{2}\right)  ^{2}$ .%

\begin{figure}
[h]
\begin{center}
\includegraphics[
height=2.0003in,
width=3in
]%
{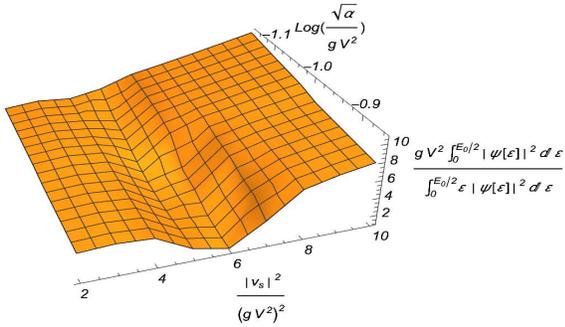}%
\caption{Dependence of the population distribution inverse width at the the
edge as function of the parameters $\alpha$ and $|v_{s}|^{2}$ for $\left\vert
E_{0}\right\vert =5gV^{2}$ and $E_{s}=2gV^{2}$ .}%
\label{FigD4}%
\end{center}
\end{figure}

In Fig.\ref{FigD4} we depict the inverse of the population distribution width
as function of the parameters $\alpha$ and $|v_{s}|^{2}$ for $\left\vert
E_{0}\right\vert =9gV^{2}$ and $E_{s}=8.7gV^{2}$. One can see that even in the
presence of a spoiler, the slower the approach rate $\alpha$ to the continuum
is, the narrower the distribution. As compared to the case of the abrupt
"switch on", one can gain narrowing by orders of magnitude.

%

\begin{figure}
[h]
\begin{center}
\includegraphics[
height=2.2572in,
width=3.3857in
]%
{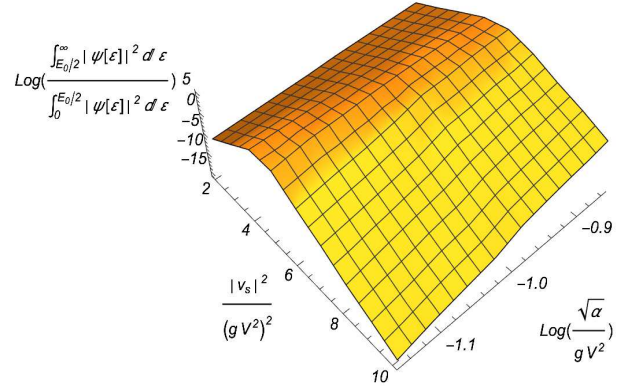}%
\caption{Ratio of the total population of the continuum edge and the total
population in the vicinity of the spoiler as a function of the spoiler
strength $|v_{s}|^{2}/\left(  gV^{2}\right)  ^{2}$ and the adiabaticity
parameter $\sqrt{\alpha}/gV^{2}$, for for $\left\vert E_{0}\right\vert
=5gV^{2}$ and $E_{s}=2gV^{2}$. The plane corresponds to equal populations at
the edge and around the spoiler.}%
\label{FigSp2}%
\end{center}
\end{figure}

\begin{figure}
[h]
\begin{center}
\includegraphics[
height=2.1318in,
width=3.1981in
]%
{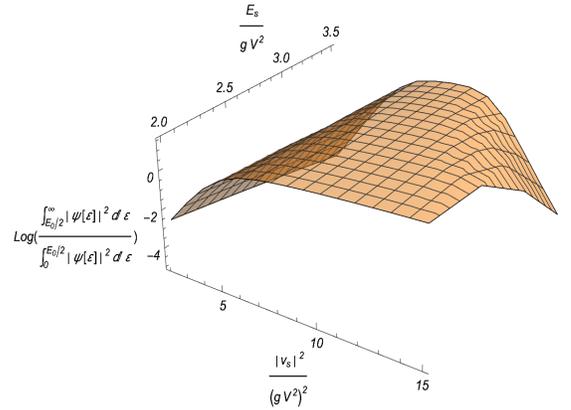}%
\caption{Ratio of the total population of the continuum edge and the total
population in the vicinity of the spoiler as a function of the spoiler
strength $|v_{s}|^{2}/\left(  gV^{2}\right)  ^{2}$ and the spoiler position
$E_{s}/gV^{2}$ for the adiabaticity parameter $\sqrt{\alpha}/gV^{2}=1$ \ and
for $\left\vert E_{0}\right\vert =5gV^{2}$ . The plane corresponds to equal
populations at the edge and around the spoiler.}%
\label{FIGSp3}%
\end{center}
\end{figure}

Answering the second question, we numerically consider the ratio of the total
population of the continuum edge and the total population in the vicinity of
the spoiler. In Fig.\ref{FigSp2} we present the result of the calculation in
function of the spoiler strength $|v_{s}|^{2}/\left(  gV^{2}\right)  ^{2}$ and
the adiabaticity parameter $\sqrt{\alpha}/gV^{2}$%
In Fig.\ref{FIGSp3} we present the dependence of the ratio between these
populations as a function of the position of the spoiler state in energy and
the spoiler strength.
One sees, that the overall population near the spoiler state gets smaller then
the approach rate $\alpha$ decreases, in complete agreement with the
Landau-Zener result for the nonadiabatic transition in a two-level system,
with the only difference, that the population does not stay in the spoiler
state itself, but gets redistributed among the band levels in its vicinity
following the profile of Fig.\ref{FigD3}.

\section{Numerous spoiler states, statistical description.}

Presence of a single spoiler state changes the energy distribution of the
population. In spite of the fact that the spoiler state by itself does not
have significant population in the long time limit, it leads to appreciable
population of the neighboring weakly coupled states of the quasicontinuum. For
the abrupt switch-on of the interaction, the populated states mainly locate on
the "blue" side, that is at the energies larger relative to the spoiler state
energy, while for the adiabatic case, the situation is opposite -- the
populated states are on the red side,as one can see in the figures above. The
natural question arises: what kind of the population distribution attains when
there are many spoiler states?

Expression (\ref{SchEqSol}) for the Fourier transform of the amplitude
(\ref{SchEqEd2}) of the quasicontinuum state $n$
\begin{equation}
\psi_{n}\left(  \varepsilon\right)  =\frac{1}{\left(  \varepsilon-E_{0}%
-\sum_{m=1}^{M}\frac{V_{m}^{2}}{\varepsilon-E_{m}}\right)  \left(
\varepsilon-E_{n}\right)  },\nonumber
\end{equation}
remains valid in this case, but the replacement of the sum by the integral no
longer can be performed. In order to understand the reason for this, let us
assume that there are many spoiler states with different sizes of couplings
$V_{k}$, that are irregularly distributed over the band with energies
$E_{m_{k}}$. The total number of the spoilers with a given $V_{k}$ is $M_{k}$.
We therefore rewrite the sum in the denominator regrouping the spoiler states
of the same size of interaction
\[
\sum_{n=1}^{M}\frac{V_{m}^{2}}{\varepsilon-E_{m}}\rightarrow\sum_{k}V_{k}%
^{2}\sum_{m_{k}=1}^{M_{k}}\frac{1}{\varepsilon-E_{m_{k}}}.
\]

The Heisenberg return time for each of the sums $\sum_{m_{k}=1}^{M_{k}}%
\frac{1}{\varepsilon-E_{m_{k}}}$ is given by the spectral density of the
spoilers of this type, and is much shorter than the Heisenberg return time of
the entire spectrum. This implies, that each group of $M_{k}$ spoiler states,
after having initially absorbed a certain amount of the population of the
ground state $0$ , returns it back, and this returned population is further
redistributed over the other states of the band. This process is neither
adiabatic nor abrupt, but occurs with a certain (rather complicated) time
dependence characterized by a typical time scale specific for the energy
spectrum $E_{m_{k}}$ of the group under consideration.

No consistent statements can be done in the general case of a complex specific
distribution of the spoiler states. Still, one can consider an average
behavior of such systems by making an assumption about statistical
distribution of the spoilers, that is tackling the problem following the
Wigner's idea of ensemble average. However, the Gaussian statistical
ensembles, traditionally employed as models of complex spectra, imply chaotic
quantum dynamics of the system with completely destroyed quantum numbers of
the parts comprising the system, which is not the case for the problem under
consideration. In fact, in the situation under consideration, each eigen state
of the quasicontinuum corresponds to the tensor product of the individual
states of two-level atoms. Therefore, we choose as model the opposite extreme
and assume, that the energy position of each spoiler state is statistically
independent of the positions of other spoilers.

Performing the statistical description, one can no longer consider the state
amplitudes, since the ensemble average results in the average of their phases
and thus completely destroys all the information about state populations. The
ensemble average has to be done for the populations, which we write in the
form%
\begin{align}
\left\langle \rho_{n}\left(  \varepsilon,\xi\right)  \right\rangle  &
=\left\langle \frac{1}{\left(  \varepsilon-E_{0}-\sum_{m=1}^{M}\frac{V_{m}%
^{2}}{\varepsilon-E_{m}}\right)  \left(  \varepsilon-E_{n}\right)  }\right.
\nonumber\\
&  \left.  \frac{1}{\left(  \xi-E_{0}-\sum_{m=1}^{M}\frac{V_{m}^{2}}{\xi
-E_{m}}\right)  \left(  \xi-E_{n}\right)  }\right\rangle , \label{AvPop}%
\end{align}
where the angular brackets denote the averaging. After the Fourier
transformation, multiplication by the factor $e^{-i\left(  \varepsilon
-\xi\right)  t}$ and integration over $\varepsilon$ and $\xi$ \ from
$-\infty\pm0i$ to $\infty\pm0i$ with the upper sign corresponding to
$\varepsilon$ and the lower to $\xi$, this expression will give the time
dependent ensemble average population of the state $n$.

In order to perform the ensemble average for the semi-infinite inhomogeneous
band, we employ the same procedure as has been earlier done \cite{Akulin} for
the infinite band, by rewriting the fractions in Eq.(\ref{AvPop}) with the
help of two auxiliary variables $\tau$ and $\theta$ as the Laplace images of
the exponents%
\begin{align}
\left\langle \rho_{n}\left(  \varepsilon,\xi\right)  \right\rangle  &
=\int\limits_{0}^{\infty}\left\langle \frac{e^{i\tau(\varepsilon-E_{0}%
-\sum_{m=1}^{M}\frac{V_{m}^{2}}{\varepsilon-E_{m}})}}{\varepsilon-E_{n}%
}\right. \nonumber\\
&  \left.  \frac{e^{-i\theta\left(  \xi-E_{0}-\sum_{m=1}^{M}\frac{V_{m}^{2}%
}{\xi-E_{m}}\right)  }}{\xi-E_{n}}\right\rangle d\tau d\theta,
\label{AvPopLap}%
\end{align}
allowing the factorization%
\begin{align}
\left\langle \rho_{n}\left(  \varepsilon,\xi\right)  \right\rangle  &
=\int\limits_{0}^{\infty}\frac{e^{i\tau(\varepsilon-E_{0})-i\theta\left(
\xi-E_{0}\right)  }}{\left(  \varepsilon-E_{n}\right)  \left(  \xi
-E_{n}\right)  }\nonumber\\
&  \prod\limits_{m=1}^{M}\left\langle e^{\frac{i\theta V_{m}^{2}}{\xi-E_{m}%
}-\frac{i\tau V_{m}^{2}}{\varepsilon-E_{m}}}\right\rangle d\tau d\theta.
\label{ProdPop}%
\end{align}

Let us concentrate now at the energy average for each individual level%
\[
\left\langle e^{i\left(  \frac{\theta V_{m}^{2}}{\xi-E_{m}}-\frac{\tau
V_{m}^{2}}{\varepsilon-E_{m}}\right)  }\right\rangle =\frac{1}{\Gamma}%
\int\limits_{0}^{\Gamma}e^{i\left(  \frac{\theta V_{m}^{2}}{\xi-E_{m}}%
-\frac{\tau V_{m}^{2}}{\varepsilon-E_{m}}\right)  }dE_{m}%
\]
randomly distributed within a broad energy band $\left(  0,\Gamma\right)  $,
with $\Gamma\rightarrow\infty$. Note, that actually, the width can be set to
infinity, it just requires to take into account some logarithmic change of the
level energy $E_{0}\rightarrow\bar{E}_{0}$, similar to the case of the
homogeneous band. The argument of the exponent tends to zero with increasing
energy, and hence the integrand is close to unity in the most part of the
integration interval. Therefore the average is very close to unity and one can
write%
\[
\left\langle e^{\frac{i\theta V_{m}^{2}}{\xi-E_{m}}-\frac{i\tau V_{m}^{2}%
}{\varepsilon-E_{m}}}\right\rangle \simeq e^{\frac{1}{\Gamma}\int
\limits_{0}^{\Gamma}\left(  e^{\frac{i\theta V_{m}^{2}}{\xi-E}-\frac{i\tau
V_{m}^{2}}{\varepsilon-E}}-1\right)  dE}.
\]

The expression obtained allows one to explicitly write the product of the
contribution of the individual levels Eq.(\ref{ProdPop}) in the form of the
sum in the exponent replaced by the integral%
\[
\prod\limits_{m=1}^{M}\left\langle e^{i\left(  \frac{\theta V_{m}^{2}}%
{\xi-E_{m}}-\frac{\tau V_{m}^{2}}{\varepsilon-E_{m}}\right)  }\right\rangle
\rightarrow e^{\int\limits_{0}^{\Gamma}dE\int dV\ g\left(  V\right)  \left(
e^{\frac{i\theta V^{2}}{\xi-E}-\frac{i\tau V^{2}}{\varepsilon-E}}-1\right)
},
\]
where $g\left(  V\right)  $ stands for the spectral density of the states of
the band levels with the coupling $V$. We finally obtain the equation%
\begin{align}
\left\langle \rho_{n}\left(  t\right)  \right\rangle  &  =V_{n}^{2}%
\int\limits_{-\infty+i0}^{\infty+i0}d\varepsilon\int\limits_{-\infty
-i0}^{\infty-i0}d\xi e^{-i\left(  \varepsilon-\xi\right)  t}\nonumber\\
&  \int\limits_{0}^{\infty}d\tau d\theta\frac{e^{i\tau(\varepsilon
-E_{0})-i\theta\left(  \xi-E_{0}\right)  +F}}{\left(  \varepsilon
-E_{n}\right)  \left(  \xi-E_{n}\right)  } \label{AvPoptime}%
\end{align}
which represents the population distribution over the band in the form of the
four-fold integral, while all the information about the behavior of the system
is contained in the function%
\begin{equation}
F\left(  \varepsilon,\xi,\theta,\tau\right)  =\int\limits_{0}^{\Gamma}dE\int
dV\ g\left(  V\right)  \left(  e^{\frac{i\theta V^{2}}{\xi-E}-\frac{i\tau
V^{2}}{\varepsilon-E}}-1\right)  . \label{FunF}%
\end{equation}

We just note that for the first, linear, term of the Taylor expansion of the
exponent over $V^{2}$, the contribution of the upper limit $\Gamma$ can be
included, as earlier, to the energy $\bar{E}_{0}$ of the isolated level, while
for all higher terms the upper integration limit can be directly set to
infinity. With this remark, we consider $\Gamma$ to be infinite hereafter,
unless the allowance of the finite size of this quantity is essential indeed.

\section{Intermediate time asymptotic for the population distribution.}

Analytic calculation of a four-fold integral Eq.(\ref{AvPoptime}) of the
general type of $F$ given by Eq.(\ref{FunF}) is not something one can easily
do. Even performing an approximate calculation with the stationary point
method, one may encounter difficulties in the case where the stationary points
are not parabolic. Therefore, our aim now is to find a regime where some
consistent statements still can be done.

We first note, that in contrast to the regime, where the couplings $V$ can be
considered as small, such that the Taylor expansion of the Eq.(\ref{FunF}) in
$V^{2}$ returns us back to the model of the uniform band considered in
Sec.\ref{Uniformband}, for the case of strongly coupled spoiler states, we
have to identify another regime, where $V^{2}$ are considered not
perturbatively. For this case, one can employ the Taylor expansion in the
population frequency instead of the couplings.

This regime turns out to be an intermediate time asymptotic, where the time is
much longer than the Grover's time, but still very short as compared to the
Heisenberg return time of the entire system. This implies that the time is
sufficiently long to ensure transition to any isolated level of the band, but
still does not allow yet to see the levels as individual and isolated. In
other words, for the regime where $Vg\gg1$, we consider the time $\hbar/V\ll
t\ll\hbar g$. Note that the parameter $Vg$ is encountered in many fields of
the quantum physics, it's square is known as the "conductance" in the
superconductivity theory, and the requirement $Vg\gg1$ is known as the
quasicontinuum existence condition in the field of laser-matter interaction.

The main idea of the calculation is as follows. Since we consider the
asymptotic of times long compared to the Heisenberg time, where the population
changes slowly, we assume that the difference $\zeta$ between variables
$\varepsilon$ and $\xi$ in Eq.(\ref{AvPoptime}), corresponding to the typical
frequencies of the population distribution time variation, to be small, and
therefore we cast the average Eq.(\ref{FunF}) in the first order Taylor series
over $\zeta$ and write%
\begin{equation}
F\left(  \varepsilon,\xi,\theta,\tau\right)  =G\left(  i\frac{\theta-\tau
}{\eta}\right)  +i\zeta\frac{\theta+\tau}{2}J\left(  i\frac{\theta-\tau}{\eta
}\right)  . \label{Cast}%
\end{equation}
Here $\eta=\frac{\varepsilon+\xi}{2}$, while the Taylor coefficients are the
functions
\begin{align*}
G\left(  i\frac{\theta-\tau}{\eta}\right)   &  =\int\limits_{0}^{\Gamma}dE\int
dV\ g\left(  V\right)  \left(  e^{\frac{i\left(  \theta-\tau\right)  V^{2}%
}{\eta-E}}-1\right) \\
J\left(  i\frac{\theta-\tau}{\eta}\right)   &  =\int\limits_{0}^{\infty}dE\int
dV\ \frac{g\left(  V\right)  V^{2}e^{\frac{i\left(  \theta-\tau\right)  V^{2}%
}{\eta-E}}}{\left(  \eta-E\right)  ^{2}},
\end{align*}
that have to be found for a specific choice of the coupling matrix elements
statistics model. Now $\zeta$ and $\eta$ are our new integration variables
replacing $\varepsilon$ and $\xi$.

Note, that the expansion Eq.(\ref{Cast}) is legitimate only for the negative
values of $\eta$ that does not lead to singularities in the exponent in
Eq.(\ref{FunF}) on the integration interval, while the positive part of $\eta$
\ axis yields a contribution decaying with time, similar to that of the
quasicontinuum infinite in both sides. In other words, in Eq.(\ref{PopUnif}),
the inhomogeneity of the band affects the first and the second contributions in
a different way, -- the first one gets shifted along the real direction of the
energy axis, while the second acquires a decreasing time dependence. We will
not consider the role of the latter here in detail, but just illustrate this
general feature in a tractable toy example presented in the next Section.

In this regime of small $\zeta$ we also note, that smallness of $\zeta$
implied by the long time asymptotic also implies a large typical size of the
sum $y=\theta+\tau$ , which therefore scales as time $t$, in contrast to a
typical size of the variable difference $x=\theta-\tau$, remaining of the
order of $1/gV^{2}$. This allows one to lift up the constrain $\left\vert
\theta-\tau\right\vert <\left\vert \theta+\tau\right\vert $ \ dictated by the
requirement $\theta,\tau>0$ and perform integration over $y$ and $x$
independently. The integration over $dy$ yields a pole at $\zeta=0$, and the
contribution of this pole to the integral over $d\zeta$ yields the time
independent density distribution in the form of a two-fold integral%

\begin{equation}
\left\langle \rho_{n}\right\rangle =\int\limits_{-\infty}^{0}d\eta
\int\limits_{-\infty}^{\infty}dx\frac{V_{n}^{2}e^{-ix\left(  E_{o}%
-\eta\right)  +G\left(  x/\eta\right)  }}{\left(  \eta-E_{n}\right)
^{2}\left(  1-J\left(  x/\eta\right)  \right)  }. \label{SpDist}%
\end{equation}

In a cases of large $G\left(  x/\eta\right)  $ where the integral
Eq.(\ref{SpDist}) can be evaluated with the saddle point method, the saddle
point for $\eta$ gives the position of a "factious" level which replaces the
displaced position $\varepsilon_{0}$ of the isolated level entering the
expression Eq.(\ref{PopUnif}) for the population distribution over the uniform
quasicontinuum, while the number $\left(  1-J\left(  x/\eta\right)  \right)
^{-1}$ in the saddle point replaces the factor $\left(  1-\frac{w}%
{\varepsilon_{0}}\right)  ^{-2}$ in that expression. If the integral is not of
a saddle point type, still, the integration gives another, more involved
population distribution profile different from that of Eq.(\ref{PopUnif}).

The result obtained gives us a population distribution over an inhomogeneous
quasicontinuum, which attains at times longer relative to the typical time
required for a Rabi transition to an isolated level of the band. This
requirement implied by the assumption that the arguments in the exponents in
integrands for $G$ and $J$ are large, such that these coefficient are not
small and indeed modify the distribution obtained earlier for the case of
uniform quasicontinuum.

The physical ground of such a modification is rather transparent. Indeed, in
contrast to the states surrounded by the neighboring levels from both sides,
the amplitudes of state at the continuum edge do not decay exponentially. We
have notice this already for the uniform quasicontinuum, -- \ the states at
the edge were responsible for the $1/t\log t$ time behavior Eq.(\ref{NonExp})
of the population. Recurrences of the population from the inhomogeneous band
to the isolated level are further redirected to quasicontinuum and keep
"feeding" these very slowly decaying states, in such a way that they finally
form at the edge a different stationary population distribution.

The structure of the function $F$ depends on the statistics of the matrix
elements, and it may give rise to various dynamic behavior, similar to that
demonstrated for the infinite quasicontinuum\cite{Akulin}. In the next Section
we focus on one of the simplest, tractable, case, which still illustrates the
main phenomena associated with recurrences of the population from the band to
the isolated level -- forming of a stationary distribution different from that
of the homogeneous quasicontinuum.

\section{An analytically tractable example.}

We consider an example where the distribution $g\left(  V\right)  $ of the
level's coupling has nonzero the second $w=\int g\left(  V\right)  V^{2}dV$
and the fourth $u=\int g\left(  V\right)  V^{4}dV$ moments, while all higher
moments are ignored. This model does not look very realistic for the case of
multiphoton transitions in ensembles of interacting Rydberg atoms, where the
distribution rather has a form $g\left(  V\right)  \propto V^{\alpha+\beta\log
V}$ , but still it allows one to immediately find an explicit expression for
$F$. Apart from the contributions of the order $w$ that have been considered
for the case of the homogeneous band, it suggests three more terms, emerging
from the second order Taylor expansion in Eq.(\ref{FunF}):
\begin{align*}
\Delta F  &  =u\int\limits_{0}^{\infty}dE\left(  \frac{i\theta}{\xi-E}%
-\frac{i\tau}{\varepsilon-E}\right)  ^{2}\\
&  =-u\frac{\theta^{2}}{\xi}-u\frac{\tau^{2}}{\varepsilon}-u2\theta\tau
\frac{\log\left(  -\varepsilon\right)  -\log(-\xi)}{\xi-\varepsilon}.
\end{align*}

We have to take into account that for positive $\varepsilon$ and $\xi$, the
logarithms have imaginary contributions $i\pi$ for $\varepsilon$ and $-i\pi$
for $\xi$, since the integration contours of the inverse Fourier
transformations for these variables are above and below the real axis,
respectively. The saddle point calculation of the integral Eq.(\ref{AvPoptime}%
) over $d\tau d\theta$ for the average density yields
\[
\left\langle \rho_{n}\left(  t\right)  \right\rangle =\int\limits_{-\infty
+i0}^{\infty+i0}d\varepsilon\int\limits_{-\infty-i0}^{\infty-i0}d\xi
\frac{V_{n}^{2}e^{-i\left(  \varepsilon-\xi\right)  t}}{\left(  \varepsilon
-E_{n}\right)  \left(  \xi-E_{n}\right)  }\frac{2\pi e^{\Phi}}{D}%
\]
with an explicit but cumbersome expression $\Phi$ in the exponent. We only
employ the zero and first order expansions for this quantity in $\zeta
=\varepsilon-\xi$ . In contrast, the denominator
\[
D=u\sqrt{\frac{1}{\varepsilon\xi}-\left(  \frac{\log(-\varepsilon)-\log(-\xi
)}{\varepsilon-\xi}\right)  ^{2}},
\]
emerging from the saddle point calculations, we consider explicitly and
notice, that it has the Taylor expansion in $\zeta$ different for the positive
and the negative $\eta$ and reads:%
\begin{align*}
D_{<}  &  =\frac{u\zeta}{3\eta^{2}}\quad\mathrm{for}\quad\eta<0\\
D_{>}  &  =\frac{4u\pi}{\zeta}\quad\mathrm{for}\quad\eta>0.
\end{align*}
The explicit expressions for the exponents read%
\begin{align*}
\Phi_{<}  &  =\frac{\eta\left(  7\eta^{2}+\overline{E}_{o}^{2}-5\overline
{E}_{o}\eta+3w^{2}-3\overline{E}_{o}w+9\eta w\right)  }{u}\\
&  +\frac{\eta w^{2}\log^{2}(-\eta)}{u}+\frac{\eta w\log(-\eta)(-2\overline
{E}_{o}+5\eta+3w)}{u}%
\end{align*}
and%
\[
\Phi_{>}=-\frac{i\zeta\left(  \pi^{2}w^{2}+(-\overline{E}_{o}+\eta+w\log
(-\eta))^{2}\right)  }{4\pi u},
\]
respectively.

The presence of the contribution $\Delta F$ modifies the results
Eq.(\ref{PopUnif}) obtained for the homogeneous quasicontinuum. The second
term in Eq.(\ref{h1}) related to the poles at the points $\varepsilon=E$ and
$\xi=E$ ( that is for $\eta>0$) with the allowance of the factor $\frac{1}%
{D}\propto\zeta$ acquires a time dependency and disappears as $1/t^{2}$ with
the time elapse. In contrast, for $\eta<0$, the contribution of the
pole$\frac{1}{D}\propto\zeta^{-1}$ gives rise to a \ new stationary profile
\[
\left\langle \rho_{E}\right\rangle =\int\limits_{-\infty}^{0}\frac{d\eta}%
{u}\frac{3\eta^{2}6\pi e^{\Phi_{<}}}{\left(  \eta-E\right)  ^{2}}.
\]

\bigskip%

\begin{figure}
[h]
\begin{center}
\includegraphics[
height=2.0003in,
width=3in
]%
{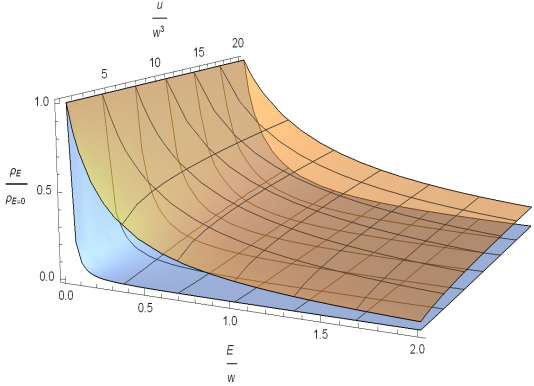}%
\caption{Intermediate asymptotic regime allowing for the recurrences of the
population from the quasicontinuum back to the isolated initially populated
level. Population distributions normalized to their values at the continuum
edge as functions of the energy position $E/w$ and the fourth moment of the
matrix elements distribution $u/w^{3}$ for two sizes of the energy gap between
the isolated level and the quasicontinuum edge. The upper surface corresponds
to $E_{o}/w=1$, and the lower one -- to $E_{o}/w=10$.}%
\label{SpSt1}%
\end{center}
\end{figure}

\begin{figure}
[h]
\begin{center}
\includegraphics[
height=2.0003in,
width=3in
]%
{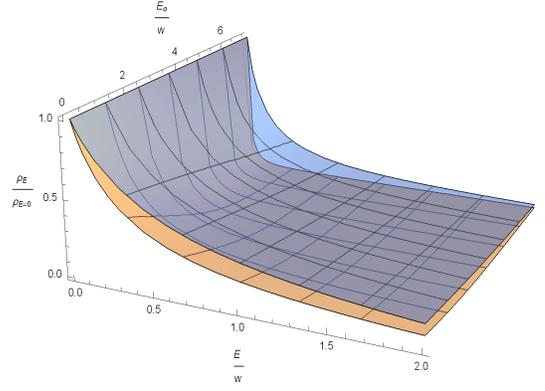}%
\caption{ Population distributions normalized to their values at the continuum
edge as functions of the energy position $E/w$ and the energy distance
$E_{o}/w$ between the isolated level and the quasicontinuum edge for two sizes
of the fourth moment of the matrix elements distribution $u/w^{3}=1$ (the
lower surface) and $u/w^{3}=10$ (the upper surface). }%
\label{SpSt2}%
\end{center}
\end{figure}

\begin{figure}
[h]
\begin{center}
\includegraphics[
height=2in,
width=3.0004in
]%
{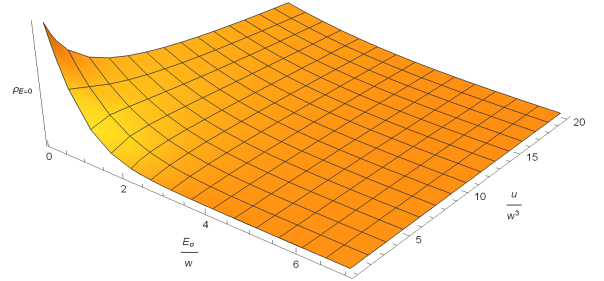}%
\caption{Population density at the quasicontinuum edge as function of the
energy gap $E_{o}/w$ and the fourth moment $u/w^{3}=1$ of the spoiler state
strength distribution. The absolute value of this quantity obtained within the
saddle point approximation cannot be considered as reliable and therefore is
not given numerically in the plot.}%
\label{SpSt3}%
\end{center}
\end{figure}

In Figs.\ref{SpSt1}-\ref{SpSt3} we depict the profiles of the population
distribution at the quasicontinuum edge suggested by this expression. In
Fig.\ref{SpSt1} one sees, that the population distribution profile gets
broader with the fourth moment $u$ increasing, but gets narrower with the
increase of the energy gap between the level and the quasicontinuum band edge,
as it is seen in Fig.\ref{SpSt2}. The last trend is opposite to the case of
the homogeneous band.%

In Fig.\ref{SpSt3} we depict the population density at the edge as function of
the gap size and the fourth moment of the coupling strength distribution.

Note that the analytically tractable case under consideration should not be
taken as a reliable prediction for the intermediate asymptotic profiles. It
just shows the mechanism of the modification of the population distribution
due to the recurrences from the inhomogeneous quasicontinuum. For the toy
example just considered, the positions of the reference points $\eta$ of the
inverse square energy distribution $\left(  \eta-E\right)  ^{-2}$ have a
rather broad distribution dictated by the function $\Phi_{<}$ emerging for
this model.

\section{Non-adiabatic transition to the edge in the presence of numerous
spoilers.}

A \ natural question arises -- how does the presence of numerous spoilers
affect the non-adiabatic transitions to the quasicontinuum from an isolated
level slowly moving toward it's edge and back? Equivalence of the
non-adiabatic transition problem and the scattering problem discussed in
Sec.\ref{NonAd} suggests to think about the problem of conductance and
scattering by disordered media that are well-studied in the Condensed Matter
Physics \cite{Efetov} with the help of the random matrix technique, which can
also be extended to the case of the absorbing media \cite{Fedorov?}.%

\begin{figure}
[h]
\begin{center}
\includegraphics[
trim=0.000000in 0.000000in 0.000000in -0.070498in,
height=3.6339in,
width=3.6115in
]%
{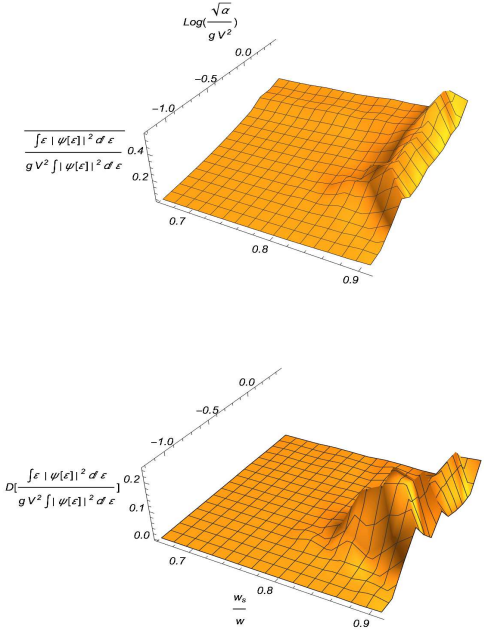}%
\caption{Width of the population distribution in the presence of the random
spoilers as function of the approach rate parameter $\alpha$ and the total
"cross-section" fraction $\left\langle g_{s}V_{s}^{2}\right\rangle /w$ of the
spoiler states. In the upper plot we present the result of averaging over 50
realization each for random positions of the spoilers that are uniformly
distributed over the band and over the couplings $V_{s}\in\left\{
0,10V\right\}  $. In the lower plot we depict the dispersion of the width.}%
\label{RandAd}%
\end{center}
\end{figure}

Technically, expressions for the states amplitudes Eq.(\ref{SchEqSol}) can be
generalized on the case of slowly moving level, -- in the Fourier
representation they adopt the operator form:
\begin{align*}
\psi_{0}\left(  \varepsilon\right)   &  =\frac{e^{-i\varepsilon t}%
}{\varepsilon-E_{0}-\alpha\frac{\partial^{2}}{\partial\varepsilon^{2}}%
-\sum_{m=1}^{M}\frac{V_{m}^{2}}{\varepsilon-E_{m}}}\psi_{t=-\infty},\\
\psi_{n}\left(  \varepsilon\right)   &  =\frac{V_{n}^{2}}{\left(
\varepsilon-E_{n}\right)  }\psi_{0}\left(  \varepsilon\right)  ,\\
\psi_{t=-\infty} &  =\lim_{t_{0}\rightarrow\infty,\beta\rightarrow0}%
\sqrt{\frac{\beta}{\pi}}\int e^{-i\varepsilon t-i\alpha\left(  t-t_{0}\right)
^{3}/3-\beta\left(  t-t_{0}\right)  ^{2}}d\varepsilon,
\end{align*}
whence expressions for the population following from their formal solution
given in terms functional integrals over some field variables are supposed to
be averaged over the statistical distribution of the band levels. The
resulting formal expression can be written explicitly. It contains functional
analogs of the average Eq.(\ref{FunF})%
\begin{equation}
F=\int dEdV\ g\left(  V\right)  \left(  e^{\frac{i\theta\left(  \xi\right)
V^{2}}{\xi-E}-\frac{i\tau\left(  \varepsilon\right)  V^{2}}{\varepsilon-E}%
}-1\right)  ,
\end{equation}
with $\theta\left(  \xi\right)  ,\tau\left(  \varepsilon\right)  $ expressed
in terms of the field variables employed.
\begin{figure}
[h]
\begin{center}
\includegraphics[
height=3.7576in,
width=3.7222in
]%
{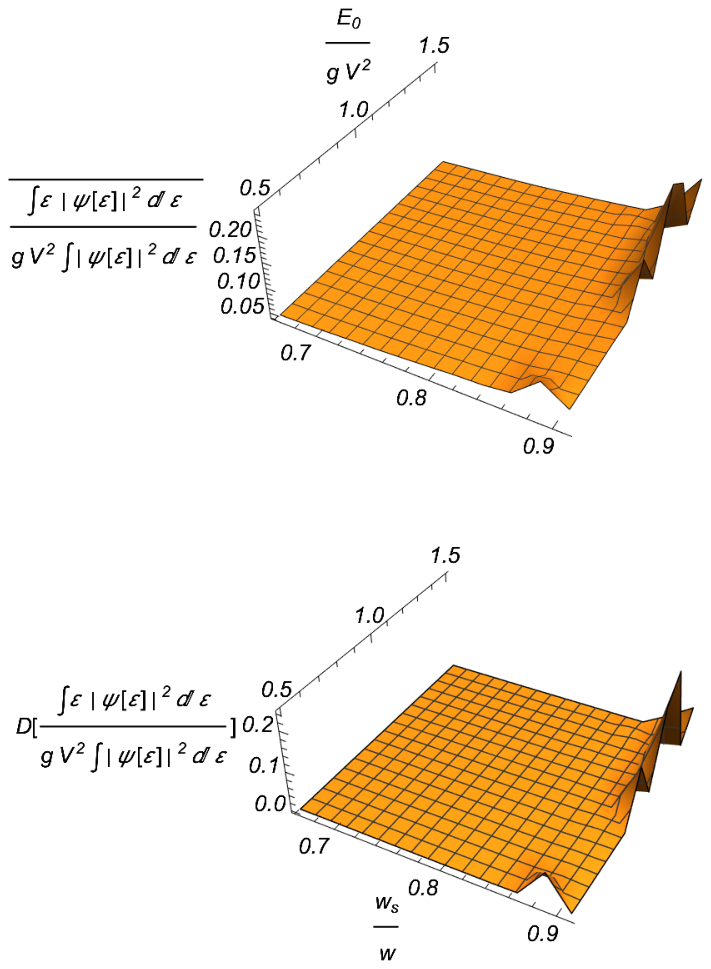}%
\caption{Width of the population distribution in the presence of the random
spoilers as function of the maximum level energy $E_{0}$ and the total
"cross-section" fraction $\left\langle g_{s}V_{s}^{2}\right\rangle /w$ of the
spoiler states. In the upper plot we present the result of averaging over 50
realization each for random positions of the spoilers that are uniformly
distributed over the band and over the couplings $V_{s}\in\left\{
0,10V\right\}  $. In the lower plot we depict the dispersion of the width
found numerically for the realizations.}%
\label{RandAdE}%
\end{center}
\end{figure}

However, unfortunately, the case under consideration does not fit well to the
models of Gaussian random potentials usually employed for description of the
disordered media that allows one to obtain explicit formula. The non-adiabatic
regime of "heavy particle", which corresponds to the optimum efficiency of the
edge population, is strongly affected by the "classically allowed" domains of
the "scattering potentials", incompatible with the model of uncorrelated
potentials implied by Gaussian statistics. We therefore leave the analytical
approach to the problem for future research, and here we just present results
of the numerical calculations performed for different random realization of
the spoiler position and strengths, each of which gives a specific population
distribution, and these distributions are averaged over the ensemble of 50
random realizations. In order to check whether or not the averaging is
meaningful, we perform averaging three times, over three sets of 50 randomly
generated realizations each.

In Figs.\ref{RandAd} and \ref{RandAdE} we present the results of these
calculations. One sees, that the distribution remains rather well localized at
the continuum edge. Still, from the dispersion of the dependencies shown in
the lower plots of each figure, one sees that the population distribution
width turns out to be rather sensitive to the particular realization of the
spoiler position and couplings in the regime of slow motion and in the regime
of large "penetration energy" $E_{0}$ of the level to the band. Presumably,
this indicates, that the main "spoiling" effect rather comes from a single,
the strongest, spoiler or from a few of the most strong of them.

\section{Summary of the results of the level-band edge dynamics.}

We summarize the results of the consideration of the population dynamics of
the quasicontinuum -- a dense band of states located in the positive part of
the energy axis $E$ with the edge at the point $E=0$. For the uniform
quasicontinuum, for the population transferred from an isolated level with the
negative energy after an abrupt switch on of the interaction, a stationary
distribution attains, which consists of two similar parts. Each of the part
can be seen as a "tail" of a $1/(E-E_{s})^{2}$ distribution, which extends to
the domain of positive energies, with the reference energies $E_{s}=E_{1}$ and
$E_{s}=0$ located at the negative part of the energy axes. Position of the
reference levels are different, -- one is close to the position of the
isolated level, which is just displaced by the Stark effect due to the
interaction with the quasicontinuum, while the other locates closer to the
continuum edge and is given by the Lambert function. This stationary
distribution attains with the time dependence $(t\log t)^{-1}$.

If the continuum is not uniform and contains "spoiler states" -- a number of
strongly coupled states locating in the quasicontinuum, the distribution gets
transformed. Each of the spoiler state, takes a certain amount of the
population and redistributes it among the other, weak, states of the
quasicontinuum that are close in energy to the spoiler. The distribution near
the edge does not experience a big change of its shape, but just loses a part
of its total population.

If the spoilers are numerous and randomly distributed in their energies and in
their coupling strengths, the overall distribution still remains localized
near the quasicontinuum edge, but the distribution profile may experience a
change of its shape, dictated by the statistics of the spoiler strengths. Also
the time law, with which the distribution attains experience a variation. This
reflects the role of the interplay of the process of population recurrences
back to the isolated level from the nonuniform quasicontinuum and the process
further redirection of this populations to the band. The resulting
distribution is a convolution of the inverse square dependence $1/(E-E_{s}%
)^{2}$ \ and the distribution of the reference energy position $E_{s}$
specific for the statistics of the spoiler strength $g(V)$.

The energy distribution of the population over the band states gets much
narrower for the case of the isolated level slowly approaching the band edge
and returning back according the law $E_{o}-\alpha t^{2}$. \ The non-adiabatic
transfer of the population to the continuum edge is \ controlled in a crucial
way by the scaled rate $\alpha/\left(  gV^{2}\right)  ^{2}$ of the level's
parabolic approach. The population profile roughly corresponds to the
dependence $\alpha/\left(  gV^{2}\right)  $ of the width on these parameters,
which directly relates to the time-energy uncertainty principle.

Presence of the spoiler states also affect this distribution, but in much
weaker way, compared to the case of the abrupt switch on of the interaction.

\section{Discussion of the results in the context of quantum computations.}

We now discuss meaning of the results obtained for the strategy of quantum
computations. We stay in the paradigm, that the quantum algorithms are
physically realized as multiphoton Raman transitions from the well defined
initial state to a narrow strip of the quantum states located near the lowest
edge of the spectrum of the corresponding Hamiltonians, and imply technically
just a set of simple manipulations such as tuning of the external fields to
required frequencies keeping coherence of the system during a sufficiently
long period along with a proper choice of the strengths of the interactions
and, if necessary, a slow variation of these parameters.

One may also want to define a parameter that can somehow characterize
"complexity" of quantum algorithms in the context of Raman excitation.
Traditional measure of complexity, that is the number of operations required
for achieving the computational goal, cannot be employed, since the Raman
excitation approach does not require a big number of operations. We therefore
propose to characterize the algorithms by the product a typical number of the
populated levels after the excitation procedure and a typical time required
for the corresponding distribution to attain. For the Grover's algorithm, this
parameter equals to the time of a single operation multiplied by the square
root of the Hilbert space dimension. In contrast, for all algorithms based on
the excitation of the quasicontinuum edge, this parameter is of the order of
the Heisenberg return time $\sim\hbar g$. In fact, for the case discussed in
Sec.\ref{Uniformband}, the typical time when distribution Eq.(\ref{PopUnif})
attains is of the order of $\hbar/w\sim\hbar/V^{2}g$, while the typical
distribution width parameter $w$ implies that the number of the populated
levels is of the order of $wg$. Same is valid for the case of Sec.\ref{NonAd},
where the typical distribution width is of the order of $\sqrt{w/\alpha}$, the
typical number of the populated states is therefore $\sqrt{w/\alpha}g$, while
the typical time $t_{typ}$ of the excitation process emerging from the
condition $\alpha t_{typ}^{2}\sim w$ also yield the same estimate of the parameter.

If one considers the process of population of just one, presumably the lowest,
state in the band, the required time simply equals the inverse gap - the
distance between the lowest and the second lowest levels. This is completely
consistent with the uncertainty principle. Of course, the same estimate
remains also valid for the Grover's algorithm when the lowest state is
separated from its nearest neighboring state by a gap of a finite size. In
other words, the time-energy uncertainty principle is the only physical
restriction of the algorithm efficiency, -- the algorithms realized as the
Raman excitation of the uniform quasicontinuum thus differ not by the
efficiency parameter but just by the "form-factor" governing the shape of the
population distribution near the edge.

The efficiency of the algorithm apparently gets worth in the presence of the
spoiler states. For an unknown structure of the distribution of the Raman
couplings over the spectrum, realization may require an optimization
procedure, that is aimed to diminish the influence of the role of "spoilers".
This means, that the excitation of the atomic ensemble has to be performed
several times in order to sequentially identify position of the spoilers
strongly affecting the population distribution, and adjusting the excitation
parameters in a way excluding this influence. The strategy in question should,
in our opinion, be the following. One starts by estimating the average value
of the interaction and the state density. Next one performs a series of
excitations and experimentally measure the populated energy eigenstates and
finds the corresponding energies. If the identified energies are far from the
edge, one can guess, that the role of the "spoiler" states is dominating, and
therefore the size of the interaction (or the detuning change velocity) has to
be reduced. This has to be done a number of times until the population of
"spoilers" became unlikely. Non-adiabatic algorithm seems to be more suitable
for such a strategy. In some sense, such a procedure is an analog of that
performed in the approach of Ref.\cite{Farhi2}, where the parameters of the
excitation protocol subject to variation in function of the atomic ensemble
average energy measurement.

\section{Conclusion}

We conclude by summarizing the concept of Quantum Algorithms seen as a Raman
excitation of an ensemble of Rydberg Atoms. Quantum computation implies
transition from a well-defined initial quantum state of an ensemble of
two-level systems to the target quantum state of a well-defined energy
position and initially unknown state vector given in the computational basis,
that is terms of binary occupation numbers of the two-level registers, whence
measurement of the register states yield the required result of the
computation. The process also implies existence of the hardware and the
software. \ I the context of ensemble of Rydberg atoms, the software is seen
as an interatomic interaction Hamiltonian diagonal in the computational basis
and an interaction Hamiltonian controlled by external laser fields, while the
target state corresponds to the edge of the spectrum. Exact calculation means
complete transfer of the population to the target state at the edge, while the
approximate calculation, in this context, mean transfer of the population to a
group of states close to the edge.

In other words, in the framework of such a concept, Quantum Algorithms are
seen as controlled processes of the population dynamics near the spectrum
edge, where the accuracy of the approximation turns to be limited \ by the
time-energy uncertainty principle exclusively. The number of the populated
levels multiplied by the time required for the population remains of the order
of the Heisenberg time given by the quantum state energy density near the
edge, which in the extreme of the single populated level is of the order of
the inverse energy gap between the edge level and its nearest neighbor.
Various population strategies differ only by a form-factor related to the
shape of the population distribution profiles.

Still, there is an important complication in achieving the uncertainty
relation limit caused by eventual heterogeneity of the band level couplings,
which we call the presence of "spoiler states" -- strongly coupled levels
located rather far from the edge. In order to avoid influence of these states,
the excitation procedure has to be performed in several steps, each of which
is supposed to identify by measurement the spoiler states close to the edge at
a given strength of the Raman coupling followed by decreasing of the coupling
strength in accordance with the measured spoiler energy position in order to
eliminate its population. Sequential application of such identification and
elimination with decreasing Raman coupling strength allows one to approach the
states at the spectrum edge.


\begin{thebibliography}{99}                                                                                               %


\bibitem {Fermi} E. Fermi, \textit{Notes on Quantum Mechanics}, Sec. 23, (
University of Chicago Press, 1960).

\bibitem {Jortner Biggs} M. Bixon and J. Jortner, \textit{Intramolecular
radiationless transitions}, J. Chem. Phys. 48, 715 (1968)

\bibitem {Fano} U. Fano, \textit{Effects of configuration interaction on
intensities and phase shifts}, Phys. Rev. 124, 1866 (1961).

\bibitem {DemkovOcherov} Yu. N. Demkov and V. I. Osherov, \textit{Stationary
and nonstationary problems in quantum mechanics that can be solved by means of
contour integration}, Sov. Phys. JETP, 26, 916 (1968).

\bibitem {Akulin}V. M. Akulin, \textit{Dynamics of Complex Quantum Systems},
(Springer, New York, 2014).



\bibitem {Farhi} E. Farhi, J. Goldstone, S. Gutmann, J. Lapan, A. Lundgren, and
D. Preda, \textit{A Quantum Adiabatic Evolution Algorithm Applied to Random
Instances of an NP-Complete Problem}, Science 292,  Issue 5516, 472 (2001); E.
Farhi, J. Goldstone, S. Gutmann, and M. Sipser, \textit{Quantum Computation by
Adiabatic Evolution}, arXiv:quant-ph/0001106 (2000).

\bibitem {Lidar} T. Albash, and D. A. Lidar, \textit{Adiabatic Quantum
Computing}, Rev. Mod. Phys. 90, 015002 (2018).



\bibitem {Farhi2} E. Farhi, J. Goldstone and S. Gutmann, \textit{A Quantum
Approximate Optimization Algorithm}, arXiv:1411.4028 (2014).



\bibitem {Lukin2} S. Ebadi et al, \textit{Quantum Optimization of Maximum
Independent Set using Rydberg Atom Arrays}, arXiv:2202.09372v1 (2022).

\bibitem {Lukin} M. D. Lukin, M. Fleischhauer, R. Cote, L. M. Duan, D. Jaksch,
J. I. Cirac, and P. Zoller, \textit{Dipole Blockade and Quantum Information
Processing in Mesoscopic Atomic Ensembles}, Phys. Rev. Lett. 87, 037901 (2001).

\bibitem {Mourachko} I. Mourachko, D. Comparat, F. de Tomasi, A. Fioretti, P.
Nosbaum, V.M. Akulin, and P. Pillet, \textit{Many-body effects in a frozen
Rydberg gas}, Phys. Rev. Lett. 80, 253 (1998).

\bibitem {Grover} L. Grover, In Proc. 28th Annual ACM Symposium on the Theory
of Computation, pages 212-219, ACM Press, New York, 1996; L. K. Grover,
\textit{Quantum mechanics helps in searching for a needle in a haystack}, Phys. Rev.
Lett. 79, 325 (1997).



\bibitem {Ioptique} A. Browaeys, and T. Lahaye, \textit{Many-body physics with
individually controlled Rydberg atoms}, Nat. Phys. 16, 132-142 (2020).

\bibitem {Zoller} A. Glaetzle, R. van Bijnen, P. Zoller, et al., \textit{A
coherent quantum annealer with Rydberg atoms}, Nat Commun. 8, 15813 (2017).

\bibitem {Lukin2} H. Pichler, S.-T. Wang, L. Zhou, S. Choi,  and M. D. Lukin,
\textit{Quantum Optimization for Maximum Independent Set Using Rydberg Atom
Arrays}, arXiv:1808.10816 (2018).

\bibitem {Efetov} K.Efetov, \textit{Supersymmetry in Disorder and Chaos,
}, (Cambridge University Press, New-York, 1997).

\bibitem {Fedorov?} S. B. Fedeli, Y. V. Fyodorov, \textit{Statistics of
off-diagonal entries of Wigner K-matrix for chaotic wave systems with
absorption, } J. Phys. A 53, 165701 (2020).
\end{thebibliography}
\end{document}